\documentclass[aps,twocolumn,twoside,secnumarabic,balancelastpage,amsmath,amssymb,hyperref=pdftex,superscriptaddress 
]{revtex4}
\usepackage{graphics}      % standard graphics specifications
\usepackage[pdftex]{graphicx}      % alternative graphics specifications
\usepackage{longtable}     % helps with long table options
\usepackage{bm}            % special 'bold-math' package
\usepackage{verbatim}
\usepackage{mathtools}
\usepackage[colorlinks=true]{hyperref}

\begin{document}

\title{Pseudo-spin Skyrmions in the Phase Diagram of Cuprate Superconductors}

\date{\today}

\author{C. Morice}
\affiliation{Institut de Physique Th\'eorique, CEA, Universit\'e Paris-Saclay, Saclay, France}
\author{D. Chakraborty}
\affiliation{Institut de Physique Th\'eorique, CEA, Universit\'e Paris-Saclay, Saclay, France}
\author{X.\ Montiel}
\affiliation{Department of Physics, Royal Holloway, University of London, Egham, Surrey, United Kingdom}
\author{C. P\'epin}
\affiliation{Institut de Physique Th\'eorique, CEA, Universit\'e Paris-Saclay, Saclay, France}

\begin{abstract}
Topological states of matter are at the root of some of the most fascinating phenomena in condensed matter physics. Here we argue that skyrmions in the pseudo-spin space related to an emerging SU(2) symmetry enlighten many mysterious properties of the pseudogap phase in under-doped cuprates. We detail the role of the SU(2) symmetry in controlling the phase diagram of the cuprates, in particular how a cascade of phase transitions explains the arising of the pseudogap, superconducting and charge modulation phases seen at low temperature. We specify the structure of the charge modulations inside the vortex core below $T_{c}$, as well as in a wide temperature region above $T_{c}$, which is a signature of the skyrmion topological structure. We argue that the underlying SU(2) symmetry is the main structure controlling the emergent complexity of excitations at the pseudogap scale $T^{*}$. The theory yields a gapping of a large part of the anti-nodal region of the Brillouin zone, along with $q=0$ phase transitions, of both nematic and loop currents characters.
\end{abstract}

\maketitle

The pseudo-gap (PG) phase in the under-doped region of cuprate superconductors remains one of the most mysterious known states of matter. First observed as a depression in the Knight shift of nuclear magnetic resonance (NMR) \cite{Alloul89,Alloul91,Warren89}, it was soon established that, for a region of intermediate dopings around $0.08<x<0.20$, part of the Fermi surface was gapped in a region close to the $\left(0,\pi\right)$ and $\left(\pi,0\right)$ points of the Brillouin zone, called anti-nodal region because of its remoteness from the point were the $d$-wave superconducting gap changes sign on the $(0,0)-(\pi,\pi)$ segment of the Brillouin zone. In this anti-nodal region, the Fermi surface was found to be ``wiped out'', and only some lines of massless quasiparticles known as Fermi arcs to be left out \cite{Campuzano98, Vishik:2012cc, Yoshida:2012kh, He14, Vishik14}.

This puzzling situation became more complex with the observation of a reconstruction of the Fermi surface by quantum oscillation and other transport measurements in the same doping region \cite{Doiron-Leyraud07, LeBoeuf07, LeBoeuf11, Laliberte11, Sebastian12, Doiron-Leyraud13, Barisic2013, Grissonnanche:2015tl}. This was attributed to the presence of incipient charge modulations with incommensurate wave vectors developing along the crystallographic axes: $\mathbf{Q}_{x},\mathbf{Q}_{y}\simeq0.3\times\left(2\pi/a\right)$, where $a$ is the lattice spacing in a tetragonal structure, detected by X-ray scattering \cite{Ghiringhelli12, Chang12, Achkar12, Blanco-Canosa13, Blackburn13a, Blackburn13b, Thampy13, Blanco-Canosa14, Tabis14, Comin14, Comin:2015vc, Comin:2015ca}. In real space, patches of charge modulation of a size of the order of twenty lattice sites have been observed at low temperatures ($T\sim4$ K) using both scanning tunneling microscopy (STM) \cite{Hoffman02,daSilvaNeto:2014vy, Mesaros:2016fe} and nuclear magnetic resonance (NMR) \cite{Wu:2015bt} measurements. These take the form of oscillations of the charge density on the copper oxide planes of a frequency comparable to twice the lattice spacing. The amplitude of these oscillations decreases away from its centerpoint in real space and disappears around ten lattice lengths away from it.

Charge modulations were observed at the core of the superconducting vortices, below the superconducting transition temperature ($T_{c}$). When voltage bias is increased, these modulations persist until the applied voltage reaches the energy scale corresponding to the formation of the pseudogap: $\Delta_{PG}$ \cite{Hamidian15,Hamidian15a}.

\begin{figure}
\centering \includegraphics[width=8.5cm]{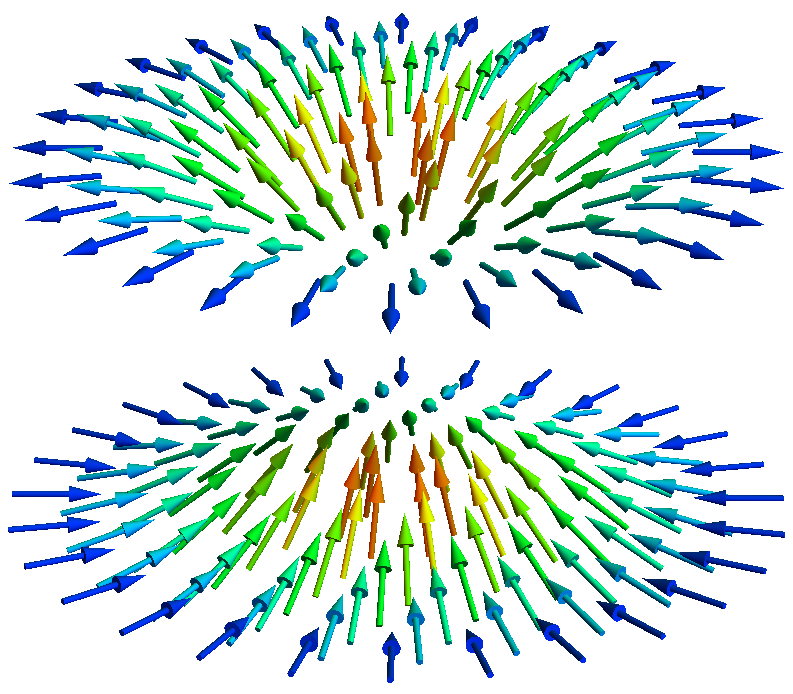} \caption{In the some regions of real space, the SU(2) order parameter is constrained to a two-dimensional hemisphere, where the vertical axis corresponds to a charge order parameter, and the horizontal plane to the superconducting order parameter. This leads to the proliferation of merons (or half-skyrmions) in meron-antimeron pairs. Note that in the center of such a meron or antimeron stands a vector with no superconducting component, and a maximal charge component.}
\label{fig:skyrmions}
\end{figure}

\begin{figure}
\centering
\includegraphics[width=8.5cm]{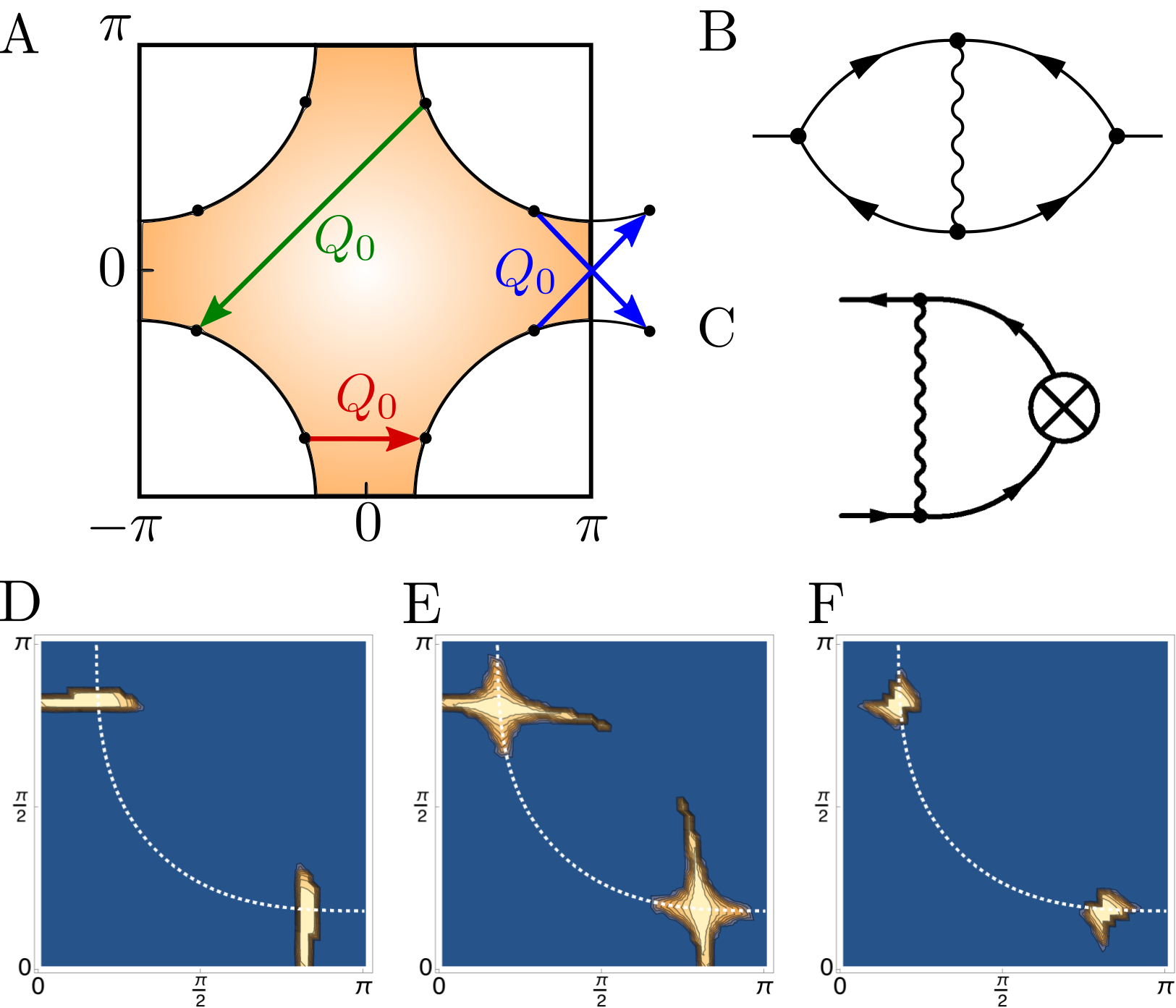}
\caption{(A): The charge modulation wave vector can take all the values connecting two hot-spots: parallel to the crystal axes (red), diagonal (blue) and the antiferromagnetic ordering vector (green), as defined in \cite{Montiel16}. (B): The SU(2) fluctuations lift the degeneracy between the various charge orders. (C): Diagram for the gap equations corresponding to the charge and superconducting order parameters. (D-F): Solving the gap equation for $\chi$ with the three ordering wave vectors shown in (A): parallel, diagonal and antiferromagnetic gives the three solutions from left to right, respectively \cite{Montiel16}.}
\label{fig:brillouin-zone} 
\end{figure}

Below the pseudogap onset temperature $T^{*}$, loop currents have been detected \cite{Fauque06, ManginThro:2014js, ManginThro:2015fg}, and the areas exhibiting charge modulations coexist with zones with long-range nematic order \cite{Mesaros:2011jj,Sato2017}, reminiscent of the vicinity of a smectic-nematic transition. The latter are more and more numerous compared to charge-modulated areas when the temperature approaches $T^{*}$ \cite{Gomes:2007ks}. Simultaneous measurements of the real and reciprocal space spectral functions however established that the opening of the pseudogap is correlated with the presence of charge modulations in real space \cite{Kohsaka07}. The whole real space picture has led to the image of an ``ineluctable complexity'' inherent to cuprate superconductors and driven by strong quantum fluctuations in the vicinity of the Mott transition in two dimensions \cite{Fradkin:2015ch}.

Here we argue that the presence of an underlying SU(2) symmetry in the under-doped region sheds light on the variety of observed phenomena and clarifies the mysteries of the real space picture. This SU(2) symmetry relates the charge and superconducting orders, similarly to the U(1) symmetry between the two components of a superconducting order parameter. First, we describe the starting short-range antiferromagnetic model and its order-by-disorder treatment, which gives rise to a pseudogap phase governed by $O(4)$ SU(2) fluctuations which stabilise $d$-wave superconducting, nematic and axial orders. Then we geometrically interpret the proliferation of local defects by introducing an SU(2) order parameter which follows naturally from the previous derivation. Finally, we describe its topological structure and the cascade of phase transitions it generates.

\section*{Short-range antiferromagnetic model}

We start by describing how an order-by-disorder treatment of a simple short-range antiferromagnetic model was shown to give rise to $d$-wave superconducting, nematic and charge orders \cite{Montiel16}.

Our starting point is that short-range antiferromagnetic interactions, strongly coupled to conduction electrons, are the main ingredient of the physics of the cuprates above $5\%$ doping. This leads to the most simple Hamiltonian:
\begin{align}
H & =\sum_{i,j,\sigma}c_{i,\sigma}^{\dagger}t_{ij}c_{j,\sigma}+J\sum_{\left\langle i,j\right\rangle }\mathbf{S}_{i}\cdot\mathbf{S}_{j}
\end{align}
where $t_{ij}$ is the hopping matrix from one site to another, $c_{\mathbf{k},\sigma}^{\dagger}$ creates an electron of momentum $\mathbf{k}$ and spin $\sigma$, $\mathbf{S}{}_{i}=\sum_{\alpha,\beta}c_{i,\alpha}^{\dagger}\sigma_{\alpha\beta}c_{i\beta}$ is the on-site spin operator and $\left\langle i,j\right\rangle $ denotes the summation over nearest neighbours.

One can perform a Hubbard-Stratonovich transformation on this Hamiltonian in order to decouple the interaction term in two channels. That is, transform the interacting term of the Hamiltonian in a sum of two terms, each corresponding to an order parameter. The first one is the usual $d$-wave superconducting channel described by
\begin{equation}
\Delta^{\dagger}=\frac{1}{2}\sum_{\mathbf{k},\sigma}d_{\mathbf{k}}c_{\mathbf{k},\sigma}^{\dagger}c_{\mathbf{-k},-\sigma}^{\dagger}
\end{equation}
The second one is the $d$-wave charge modulations channel at momentum $\mathbf{Q}_{0}$, described by
\begin{equation}
\chi=\frac{1}{2}\sum_{\mathbf{k},\sigma}d_{\mathbf{k}}c_{-\mathbf{k}+\mathbf{Q}_{0}}^{\dagger}c_{\mathbf{-k}}
\end{equation}
where $d_{\mathbf{k}}= 2 \cos\left(2\theta_{\mathbf{k}}\right)$ is the $d$-wave factor, and $\theta_{\mathbf{k}}$ the angle spanning the Brillouin zone. It corresponds to a charge density wave order with an ordering wave vector $\mathbf{Q}_{0}$, and a $d$-wave modulation, meaning in particular that its gap exhibits a $d$-wave modulation in momentum space. The charge modulation wave vectors, shown in Fig.\ \ref{fig:brillouin-zone}A, are typically incommensurate, and taken either parallel to the crystal axes \cite{Montiel16} or diagonal \cite{Metlitski10,Efetov13}. Indeed, it can take all the values connecting two hot-spots, which are the points where the Fermi surface crosses the line where the antiferromagnetic fluctuations diverge: $\mathbf{Q}_{0}=\left(0,0\right);\left(\pm Q_{x},0\right);\left(0,\pm Q_{y}\right);\left(\pm Q{}_{x},\pm Q_{y}\right);\left(\pm\pi,\pm\pi\right)$ (Fig.\ \ref{fig:brillouin-zone}A).

\begin{figure*}
\centering \includegraphics[width=18cm]{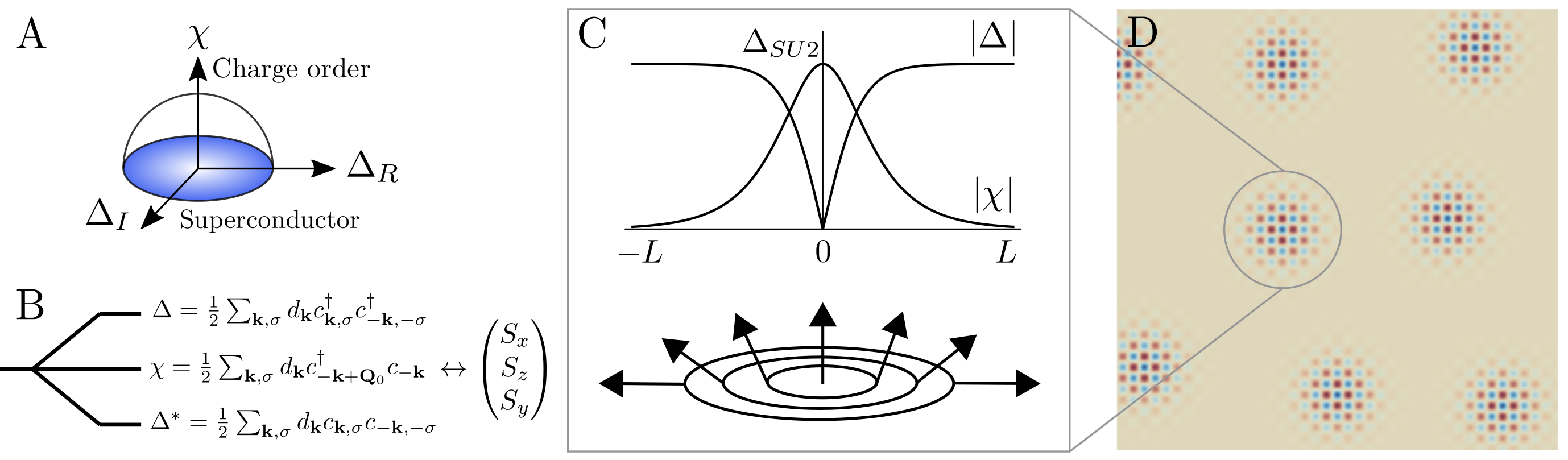} \caption{In some regions of real space, the SU(2) order parameter is constrained to two-dimensional hemispheres (A). It can therefore be mapped on a pseudo-spin vector (B). This constraint causes the proliferation of pseudo-spin merons of size L (C), which have a maximal charge component and a zero superconducting component at the core (C). The charge component of a set of these merons is schematically represented in (D), which matches STM experimental observations \cite{Hoffman02}.}
\label{fig:energy-splitting} 
\end{figure*}

This decoupling yields various possible order parameters which are all degenerate in magnitude at the Fermi surface hot-spots in the strong coupling limit, i.e.\ for $J$ much larger than the energy of the bottom of the electronic band in the anti-nodal region, as depicted in Fig.\ \ref{fig:brillouin-zone}D-F \cite{Montiel16}. Note that one could also consider decoupling the interaction term in the spin-sector in the antiferromagnetic channel, in particular for the characterization of the low-doping properties of this model. It is however likely that the antiferromagnetic fluctuations would be gapped by the emergent orders which also compete for the hot-spots.

All the previously cited possible orders are $d$-wave symmetric. This is dictated by the fact that antiferromagnetic interactions correspond to a $\mathbf{Q}=(\pi,\pi)$ wave vector. Therefore when writing the gap equation corresponding to the ladder diagram in Fig.\ \ref{fig:brillouin-zone}C, it couples adjacent anti-nodal regions of the Brillouin zone:
\begin{align}
\Theta_{\mathbf{k}} & =-T\sum_{\omega,\mathbf{q}}J_{\overline{\mathbf{q}}}\frac{\Theta_{\mathbf{k}+\overline{\mathbf{q}}}}{\Theta_{\mathbf{k}+\overline{\mathbf{q}}}^{2}+\xi_{\mathbf{k}+\overline{\mathbf{q}}}^{2}+\left(\epsilon+\omega\right)^{2}},
\label{eq:gap}
\end{align}
with $\Theta=\chi,\Delta$, $\xi_{\mathbf{k}}$ the electronic dispersion and $\epsilon$ and $\omega$ the fermionic and bosonic Matsubara frequencies and $\overline{\mathbf{q}}=q+\mathbf{Q}$. The solutions of this equation are plotted in Fig.\ \ref{fig:brillouin-zone}D-F for $\chi$ with three different choices for $\textbf{Q}_0$. In order to stabilise the gap equation there needs to be a change of sign between two such adjacent regions leading to $\Theta_{\mathbf{k}}=-\Theta_{\mathbf{k}+\mathbf{\overline{\mathbf{q}}}}$, the simplest case of which is a $d$-wave order \cite{Kloss15}.

The approach presented here has two specificities: (i) we consider a strong coupling regime where $J$ is larger than the bottom of the band in the anti-nodal region, which yields ``hot-regions" instead of hot-spots. (ii) we consider the whole frequency dependence in the gap equation \eqref{eq:gap}. $J$ appears as a weight in the frequency sum of Eq. \eqref{eq:gap} and therefore corresponds effectively to a bosonic frequency cutoff.

A similar decoupling has been used on the spin-fermion model in the vicinity of an antiferromagnetic quantum critical point to obtain a model simpler than the one presented here \cite{Efetov13}. It is named the eight hot-spots model because the conduction electrons are only considered at the hot-spots, and the electronic dispersions are linearised close to the Fermi surface. In this eight hot-spots model, the fact that there is an instability in the charge modulation channel has been the subject of intense discussion. Indeed, if the charge modulation wave-vector $\textbf{Q}_0$ is diagonal, the Fermi surface is said to be nested, and $\textbf{k}$ and $\textbf{k}+\textbf{Q}_0$ are parallel, at least to first approximation. In this case, the gap equation \eqref{eq:gap} features two poles in different half-planes and performing the sum yields a Cooper logarithm and therefore a charge instability. This divergence disappears when the curvature of the Fermi surface is included, except in the case of superconductivity where the logarithm is independent of the curvature. If $\textbf{Q}_0$ is parallel, the Fermi surface is said to be anti-nested, and $\textbf{k}$ and $\textbf{k}+\textbf{Q}_0$ are anti-parallel. In this case, the gap equation features a double pole, and performing the sum gives zero, and therefore no instability \cite{Mishra2015}. However, it was found that this is not true if we consider the frequency dependence of the gap equation, or if we go to second order in the interaction, where there is a back-and-forth scattering between $\textbf{k}$ and $\textbf{k}+\textbf{Q}_0$ \cite{Wang14,WangReply}. Therefore, in the weak coupling regime, the charge instability only arises if $\textbf{Q}_0$ is diagonal. One needs to be away from weak coupling to see an instability for a parallel wave-vector. Here we are in the strong coupling regime, and therefore we obtain an instability even if $\textbf{Q}_0$ is parallel.

\section*{Order-by-disorder treatment}

It has been shown that the charge and superconducting order parameters $\chi$ and $\Delta$ are related by an SU(2) symmetry in a region of the Brillouin zone, in the sense that one can define an SU(2) algebra relating the two \cite{Montiel16}. This symmetry is exact on a line of the Brillouin zone joining the hot-spots, and is broken away from it \cite{Montiel16}. This naturally causes the arising of the fluctuations associated to this symmetry, which we call SU(2) fluctuations \cite{Montiel16}.

The degeneracy of the various channels at the hot-spots introduced above has been shown to be lifted by considering the SU(2) fluctuations through the diagram in Fig. \ref{fig:brillouin-zone}B \cite{Montiel16}, similarly to what happens in the \emph{order-by-disorder} formalism, first described by Villain in the classical context \cite{Villain80}.

Remarkably, the choice of the starting charge modulation wave vector becomes irrelevant at this point, since it was found that the SU(2) fluctuations select three wave vectors characterizing respectively $d$-wave nematic ordering at $\mathbf{Q}_{0}=\left(0,0\right)$ and axial modulations with or without $C_{4}$ symmetry breaking at $\mathbf{Q}_{0}=\left(\pm Q_{x},0\right)$ and $\left(0,\pm Q_{y}\right)$ \cite{Montiel16} (Fig.\ \ref{fig:brillouin-zone}A). Both nematic and axial orders are therefore naturally selected by the SU(2) fluctuations.

In order to define the set of operators relating $\Delta$ and $\chi$, which form the SU(2) algebra, we use the order parameter $\chi=\frac{1}{2}\sum_{\mathbf{k},\sigma} \overline{d}_{\mathbf{k}}c_{\overline{\mathbf{k}}}^{\dagger}c_{\mathbf{-k}}$ with $\overline{\mathbf{k}}=-\mathbf{k}+\mathbf{Q}_{0}$ and $\overline{d}_{\mathbf{k}} = (d_{\mathbf{k}} + d_{\overline{\mathbf{k}}})/2$. The operation $\overline{\mathbf{k}}$ is constrained by the closure condition of the SU(2) algebra to satisfy $\overline{\overline{\mathbf{k}}}=\mathbf{k}$ and $\overline{(-\mathbf{k})} = -\overline{\mathbf{k}}$ \cite{Montiel16}. This causes $\mathbf{Q}_{0}$ to be $\textbf{k}$-dependent, that is, this causes the charge modulations to be multi-$\textbf{k}$ \cite{Montiel16}. The lack of $\textbf{k}$-dependence would cause the arising of harmonics at multiples of $\mathbf{Q}_{0}$ \cite{Montiel16}. This $\mathbf{Q}_{0}(\textbf{k})$ can be chosen in several ways, for example by dividing the Brillouin zone in quadrants and assigning a single vector per quadrant, as mentioned implicitely above \cite{Efetov13, Montiel16}. It can also correspond to $2\textbf{k}_F$ charge modulations responsible for both the gapping of the anti-nodal region of the Brillouin zone and the formation of excitonic patches, as studied in \cite{Montiel16}.

These excitonic patches have been shown to proliferate in real space in some regions of the phase diagram \cite{Montiel16}. In the following section, we give a topological interpretation of the proliferation of local objects in real space, by introducing the SU(2) order parameter, which enables us to encompass many aspects of the phase diagram of the cuprates in an integrated manner.

\section*{SU(2) order parameter}

The order parameter that naturally emerges from the previous discussion to describe the pseudogap is a composite of $\Delta$ and $\chi$, which can be cast into the form: 
\begin{align}
\hat{\Delta}_{SU2} & =\left(\begin{array}{cc}
\chi & \Delta\\
-\Delta^{*} & \chi^{*}
\end{array}\right),\label{eq:1}
\end{align}
where $\Delta_{SU2}^{2}=\left|\chi\right|^{2}+\left|\Delta\right|^{2}$,
which is the constraint enforcing the SU(2) symmetry. Since $\chi$
and $\Delta$ are complex fields, this constraint can be written as:
\begin{align}
\Delta_{SU2}^{2} & =\chi_{R}^{2}+\chi_{I}^{2}+\Delta_{R}^{2}+\Delta_{I}^{2}.\label{eq:2}
\end{align}
where the indices $R$ and $I$ denote the real and imaginary parts of the operators,
respectively. In this picture, $\Delta_{SU2}$ represents the energy
scale below which the fluctuations between the two fields $\chi$
and $\Delta$ are dominant; this scale is thus doping dependent. Notice
that, by construction, this composite SU(2) order parameter is non-abelian.

At every doping $x$, equation
\eqref{eq:2} describes a three dimensional hypersphere $\mathbb{S}^{3}$
in a four-dimensional space. The transverse fluctuations of the order parameter on this hypersphere are naturally described by an O(4) non-linear
$\sigma$-model \cite{Efetov13} 
\begin{align}
S & =\int d^{2}x\sum_{\alpha=1,4}\frac{1}{2}\left[\frac{\rho}{T}\left(\partial_{\mu}n_{\alpha}\right)^{2}+\sum_{\alpha}m_{\alpha}n_{\alpha}^{2}\right]
\end{align}
where $\alpha=1,4$ are the four-component vector subject to the constraint
$\mathbf{n}^{2}=1$, with $n_{1,2}=\overline{\chi}_{I},\overline{\chi}_{R}$,
$n_{3,4}=\overline{\Delta}_{I},\overline{\Delta}_{R}$, where $\overline{\chi}=\chi/\Delta_{SU2}$, $\overline{\Delta}=\Delta/\Delta_{SU2}$ and the sign of the masses $m_{\alpha}$ depends on the presence or absence of an applied magnetic field. The amplitude modes, or massive modes, can be safely neglected since the energy difference between the charge and superconducting states is much smaller than both their energies.

In the specific context of the $\mathbb{S}_{3}$ sphere, no topological defect is generated, since a careful examination of the corresponding homotopy class gives $\pi_{2}\left(\mathbb{S}^{3}\right)=0$ \cite{Mermin:1979io}. In the following, we discuss the case where one degree of freedom is lost, allowing for topological defects to appear.

\section*{Topological defects}

\begin{figure}
\centering
\includegraphics[width=8.5cm]{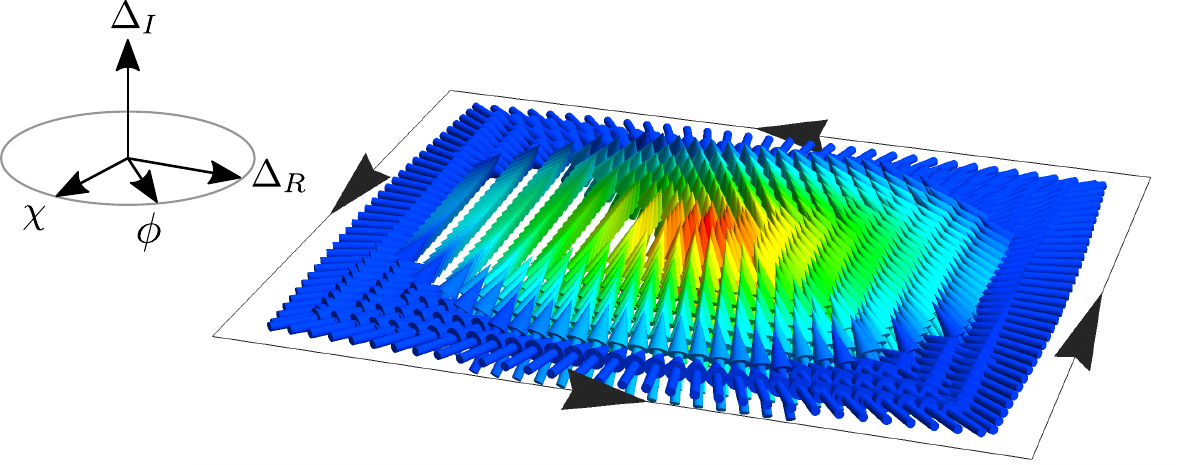}
\caption{Peculiar SU(2) meron, rotated with respect to the simple case of Fig.\ \ref{fig:skyrmions}. Here, the order parameter is purely parallel to $\chi$ at two edge points, and purely parallel to $\Delta_R$ at two other edge points. This causes the arising of a peculiar evolution of the charge order in real space, along with a superconducting current along a central axis, corresponding to a gradient of the phase of $\Delta$. Moreover, defining the order parameter $\phi = \Delta_R + i \chi$ yields a gradient of its phase along the edge, and therefore an associated current.}
\label{fig:device} 
\end{figure}

We now argue that, as the temperature is lowered, the phase of the charge modulations is frozen in some real space regions, as measured by phase-resolved STM \cite{Mesaros:2011jj, Mesaros:2016fe}. This reduces this phase to a few integer values $\pm i\pi/n$, with $n$ an integer and thus reduces the fluctuation space from $\mathbb{S}^{3}$ to a set of halves of $\mathbb{S}^{2}$, indexed by $\mathbb{Z}_{2n}$ \cite{Lee:1979hu}. These regions are thus characterised by $\Delta_{SU2}^{2}=\chi^{2}+\Delta_{R}^{2}+\Delta_{I}^{2}$, and the effective non-linear $\sigma$-model becomes $O(3)$. The space of the fluctuations is depicted in Fig.\ \ref{fig:energy-splitting}A where a fluctuating hemi-sphere is shown; $\mathbb{Z}_{2n}$ has been reduced to $\mathbb{Z}_{2}$ for the clarity of the representation, with phase $+1$ and $-1$ corresponding to the upper and lower hemi-spheres, respectively.

The second homotopy class of the $\mathbb{S}^{2}$ sphere is $\pi_{2}\left(\mathbb{S}^{2}\right)=\mathbb{Z}$, which yields the spontaneous generation of skyrmions \cite{Mermin:1979io}. In our case, in the $O(3)$ regions of real space, $\Delta_{SU2}^{2}$ fluctuates on a hemisphere and the boundary conditions corresponding to superconducting vortices give us half-skyrmions bearing a half-integer topological charge \cite{Lee:2001ib,Lee06, Sheehy1998, Goldbart1998}:
\begin{equation}
Q_{top} = \frac{1}{4 \pi} \int \textbf{n} \cdot \partial_x \textbf{n} \times \partial_y \textbf{n} \; dx dy
\end{equation}
They are also called merons and correspond to a variation of the vector $\mathbf{n}$ over one hemisphere, as illustrated in Fig.\ \ref{fig:skyrmions}, \ref{fig:energy-splitting}A and \ref{fig:energy-splitting}C. They take two equivalent typical forms, of an edgehog and a vortex, and the proliferation pattern involves meron/anti-meron pairs such as the one presented in Fig.\ \ref{fig:skyrmions}. Note that, contrary to the magnetic skyrmions observed in magnetic systems (see e.g.\ \cite{Muehlbauer:2009bc,Romming636}), here the topological structure acts on the pseudo-spin sector, with the three quantization axes $\left(S_{x},S_{y},S_{z}\right)$ corresponding respectively to $\left(\Delta_{R},\Delta_{I},\chi\right)$ (Fig.\ \ref{fig:energy-splitting}B). The choice of the quantization axis $z$ to be parallel to the charge modulation parameter $\chi$ is arbitrary but convenient, since the superconducting phase then corresponds to a simple easy plane situation (Fig.\ \ref{fig:energy-splitting}C).

Non-zero topological numbers are associated with the arising of edge currents. One can therefore imagine isolating one topological defect in order to examine these currents. In the most simple case of a single meron, such as displayed in Fig.\ \ref{fig:skyrmions}, the SU(2) order parameter along the edge is purely in the superconducting plane, with a superconducting phase varying by $2\pi$ when going around the full edge, exactly like in the case of a superconducting vortex. One can however think of a different situation, such as the particular meron depicted in Fig.\ \ref{fig:device}. It corresponds to a rotation of the axes of the simple case considered in Fig.\ \ref{fig:skyrmions} and \ref{fig:energy-splitting} such that $\Delta_I$ would now be along $z$. Note that this would mean that the phase of the charge order parameter changes along a line dividing the meron in two, on which it has zero magnitude. Experimentally, this would give rise to a supercurrent along this line, as well as a peculiar charge pattern, measureable for example via STM. Moreover, one can define a specific order parameter along the edge as $\phi = \Delta_R + i \chi$. The phase of this order parameter then rotates by $2 \pi$ when going round the sample, giving rise to a finite phase gradient, and hence to a peculiar type of current.

The mechanism that allows one to see these topological defects is the freezing of the phase of the charge order parameter component of the SU(2) order parameter. This is caused by pinning to local defects or superconducting vortices and coupling of the charge order parameter to the lattice. In this formalism there is therefore coexistence, in the pseudogap phase, of two types of regions in real space: regions where the phase of $\chi$ is continuous, and which exhibit a nematic response, and regions where the phase of $\chi$ is quantised, where merons proliferate.

Measurements of superconducting vortices below $T_{c}$ found that they bear a very specific structure where charge modulations are observed at the core \cite{Wu13a}. This corresponds to a pseudo-spin meron in whose centre the pseudo-spin vector is oriented along the $z$-axis, producing charge modulations while the superconductivity order parameter vanishes, as detailed in Fig.\ \ref{fig:energy-splitting}C. The energy associated to the creation of this vortex is intrinsically of the order of the energy splitting between the superconducting and charge modulation orders, which is precisely the typical energy scale of the superconducting coherence $\Delta_{c}\sim\frac{1}{2}k_{B}T_{c}$. Hence pseudo-spin merons will proliferate around $T_{c}$ in the under-doped region of the phase diagram, acting as a Kosterlitz-Thouless (KT) transition towards the pseudogap state \cite{Wang:2002ke,Alloul:2010ko}. Here, this proliferation is the driving mechanism behind the transition from the superconducting state to the pseudogap state, and it will be strongest close to $T_{c}$. Note that the system is phase-coherent at low temperature and is driven by this transition to the phase-incoherent pseudogap phase.

The size of the merons can be obtained from similar considerations. Indeed, if one only consider one mass for each field (setting, e.g., $m_1=m_2$ and $m_3=m_4$), the energy associated with this topological defect is $|m_1 - m_3|$, which gives us the size of the meron: $L=\hbar v_F / \sqrt{|m_1 - m_3|}$, written in Fig.\ \ref{fig:energy-splitting}C. Note that if superconductivity dominates, $\Delta_{c} = |m_1 - m_3|$.

Note that such a pseudospin analogy has also been derived in the case where the three components of the pseudospin are the density fluctuations, the ampliton and the phason operators. The dynamics of this model were studied in analogy to the spin transfer torque effect in magnetic systems and it was found that a similar non-equilibrium superconducting effect could be induced by gradients such as electric or heat gradients \cite{Garate2013}.

We have introduced the general framework of the SU(2) theory, described how SU(2) fluctuations stabilise a pseudogap state, and how some regions in real space can freeze the charge modulation phase, causing the proliferation of pseudo-spin merons. We now turn to the consequences of this formalism on experimental observations. We start by discussing the simultaneous arising of nematic and loop current orders at $T^*$, then we proceed to charge modulations in real space under applied electric field, and finally we examine the phase diagram under applied magnetic field.

\begin{figure*}
\centering \includegraphics[width=18cm]{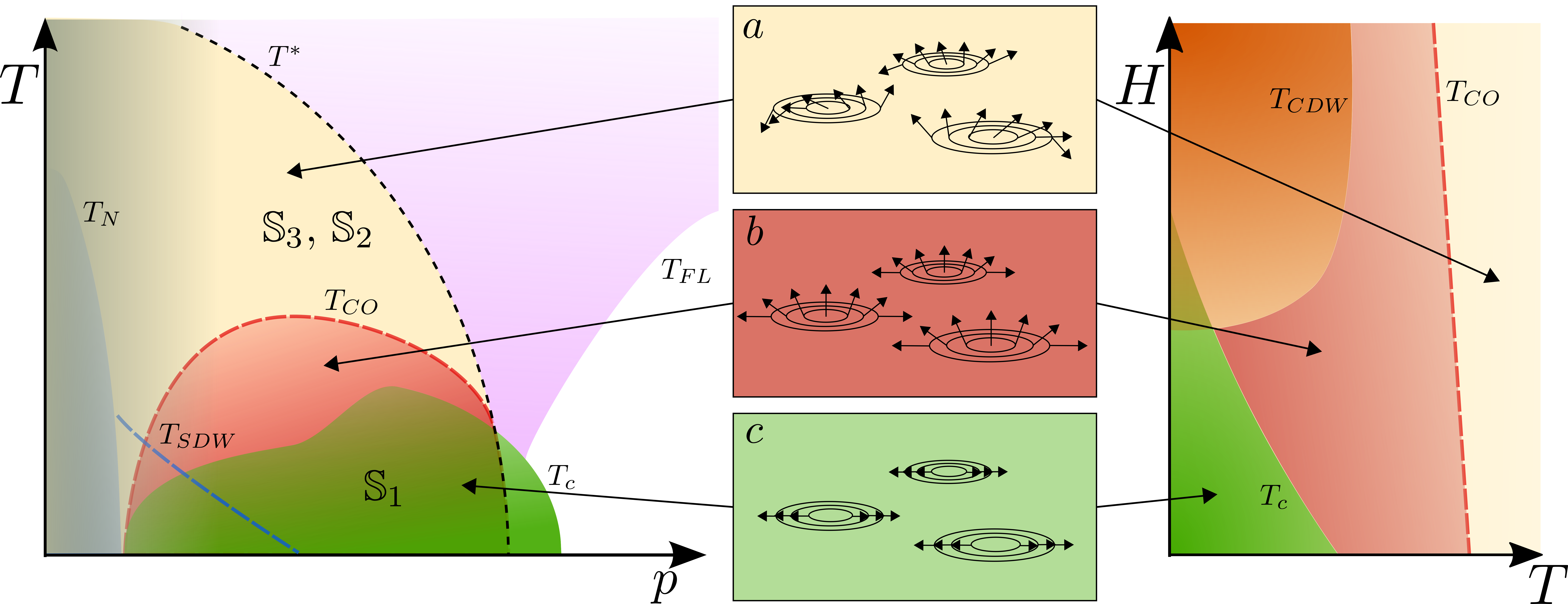} \caption{Schematic phase diagram of the cuprate superconductors in the temperature-doping plane (left) and the applied magnetic field-temperature plane at low doping (right). It features three different structures of the SU(2) order parameter field, illustrated in the middle: the pseudogap phase (a), the charge ordered phase (b) and the superconducting phase (c). The labels corresponds to the Néel temperature $T_N$, the spin density wave onset temperature $T_{SDW}$, the pseudogap onset temperature $T^*$, the charge order temperature $T_{CO}$, the superconducting transition temperature $T_c$, the Fermi liquid crossover temperature $T_{FL}$, and the charge density wave transition temperature $T_{CDW}$. The greyed area illustrates the fact that our approximation does not allow us to replicate the low doping phase diagram. Note that $T^*$ is not drawn in the right hand side phase diagram because it sits at too high temperatures.}
\label{fig:phase-diagram}
\end{figure*}

\section*{Multiplicity of orders at $T^{*}$: an ineluctable complexity?}

It is a longstanding issue whether the pseudogap temperature $T^{*}$, sketched in Fig.\ \ref{fig:phase-diagram}, corresponds to a phase transition or a cross-over. Solving this issue is made harder by the fact that the cause of the gapping of the Fermi surface also generates ``collateral'' orders which typically break discrete symmetries. Disintricating this cause and its collateral orders remains an outstanding problem.

Recently, the pseudogap line has been associated with the presence of $q=0$ orders in the form of intra unit cell orbital currents \cite{Fauque06, ManginThro:2014js, ManginThro:2015fg}, and to the breaking of the $C_{4}$ rotational symmetry leading typically to a nematic signal \cite{CyrChoiniere:2015dk}. Ultra-sound experimental data following the pseudogap line have been strengthening the case for a phase transition \cite{Shekhter13}. Although $q=0$ orders cannot open a gap in the Fermi surface, they can induce time-reversal symmetry breaking, as in the case of loop currents, or $C_{4}$ symmetry breaking, as in the case of nematicity.

The SU(2) model is in very good posture to bring an ample clarification to the situation. Indeed it does predict the simultaneous gapping of the Fermi surface in the anti-nodal region and enhancement of the nematic susceptibility at $T^{*}$, both emerging from the SU(2) fluctuations. Nematicity can then be stabilised as a collateral order below $T^{*}$, for example by a small amount of $C_{4}$ symmetry breaking due to the oxygen chains in YBCO \cite{Orth2017}. One can remark, interestingly, that the $q=0$ breaking of the $C_{4}$-symmetry is not necessarily in competition with the charge modulations.

We now turn to the possible generation of loop current orders by the SU(2) fluctuations. For this, we rely on symmetry considerations, following reference \cite{Agterberg:2014wf}. The SU(2) symmetry involves strong coupling between $\chi$ and $\Delta$, which leads us to consider the scalar field $\phi_{\textbf{Q}_{0}}=\chi_{\textbf{Q}_{0}}\Delta$, ($\phi_{\textbf{Q}_{0}}^{*}=\chi_{\textbf{Q}_{0}}\Delta^{*}$). We consider the influence of both time-reversal ($\mathcal{T}$) and parity ($\mathcal{P}$) transformations on this order parameter (see Supplemental Material for details):
\begin{equation}
\phi_{\textbf{Q}_{0}} \xrightarrow[]{\mathcal{T}} \phi_{-\textbf{Q}_{0}}^{*}, \quad \phi_{\textbf{Q}_{0}} \xrightarrow[]{\mathcal{P}} \phi_{-\textbf{Q}_{0}}
\end{equation}
We now form a loop current order parameter out of the field $\phi_{\textbf{Q}_{0}}$:
$l=\left|\phi_{\textbf{Q}_{0}}\right|^{2}-\left|\phi_{-\textbf{Q}_{0}}\right|^{2}$.
Strikingly, this new order parameter transforms as (see Supplemental Material for details):
\begin{equation}
l \xrightarrow[]{\mathcal{T}} -l, \quad l \xrightarrow[]{\mathcal{P}} -l, \quad l \xrightarrow[]{\mathcal{TP}} l
\end{equation}
which is precisely the signature of the loop current observed experimentally. Following the argument in \cite{Agterberg:2014wf}, the field $\phi_{\textbf{Q}_{0}}$ has the same symmetries as a pair density wave (PDW) order parameter, and by symmetry such a field can sustain loop currents described by $l$. These are nil outside the pseudogap phase, since $|\Delta_{SU2}|=0$. But inside the pseudogap phase, $|\Delta_{SU2}| \neq 0$ hence both $\chi$ and $\Delta$ are finite, except when $\Delta_{SU2}$ is either in the superconducting state or in the charge ordered state. The detailed study of the Ginzburg-Landau formalism for this field theory will be clarified elsewhere, and follow closely the study of the PDW detailed in reference \cite{Agterberg:2014wf}.

\section*{Charge modulations in real space}

Charge modulations (CM) were observed up to the pseudogap energy scale by STM experiments under applied electric field \cite{Mesaros:2011jj,Hamidian15a}. Bulk X-ray probes and NMR, however, reported the presence of charge-modulated areas over a doping dome which doesn't follow $T^{*}$ but decreases with $T_{c}$ as doping decreases \cite{Ghiringhelli12, Blackburn13a, Comin14}. Here lies a remarkable discrepancy : are the CM associated to the pseudogap energy scale $\Delta_{PG}$, or to the superconducting coherence energy scale $\Delta_{c}\sim\frac{1}{2}k_{B}T_{c}$?

To answer this question we first notice that the STM experiments are done at low temperature inside the superconducting phase ($T=4$ K) whereas the X-ray probes can directly reach $T^{*}$. The SU(2) scenario provides a simple explanation for this unusual situation. Indeed, in the SU(2) picture, vortices inside the ordered superconducting phase have a different structure than in standard superconductors where the normal state is a Landau metallic state. Fig.\ \ref{fig:energy-splitting}C and \ref{fig:energy-splitting}D show that the charge order coordinate of the SU(2) order parameter is finite at the center of the vortex core. This is similar to the case of SU(2) rotations between $d$-wave superconductivity and a $\pi$-flux phase \cite{Lee:2001ib,Einenkel14}. Here the SU(2) symmetry constrains charge modulations to be present inside the vortex core, as it has been actually seen by STM and NMR \cite{Wu13a}. In terms of topological defects, it is as if a meron was sitting at the center of the vortex, with the quantization axis $z$ locked to the direction of the charge modulations (Fig.\ \ref{fig:energy-splitting}C). The emergence of a finite charge order parameter in the center of superconducting vortices is caused by the existence of this third degree of freedom of the SU(2) order parameter. This gives a natural explanation to the observation by bulk probes of charge modulations in the region of the phase diagram where vortices are present, below $T_{c}$ and within a dome above $T_{c}$ \cite{Wang:2002ke}.

At the operation temperature of STM experiments ($T=4$ K), the quantization axis of the merons is still locked to the charge modulation direction, up to the energy scale $\Delta_{SU2}\simeq\Delta_{PG}$, typical of the pseudogap. As the temperature is raised above the superconducting fluctuations dome, the distribution of quantisation axes across merons becomes fully random, as depicted in Fig.\ \ref{fig:phase-diagram}, and the CM are impossible to observe. Since the SU(2) theory features the coexistence of $O(3)$ and $O(4)$ real space regions, the first of which becoming less and less numerous when the temperature is raised, we can see that there are less and less CM as we get closer to $T^{*}$, which elucidates experimental
data \cite{Kohsaka:2008bz,Gomes:2007ks}. This disappearance of the CM patches when raising the temperature close to $T^*$ was previously explained by dislocations of the phase of the charge order, interpreted in the framework of a nematic/smectic transition where the nematic order competes with the CM patches \cite{Mesaros:2011jj}. Note that in our case, there is no competition between the two, and that the disappearance of the topological defects when temperature is raised, as $O(3)$ regions make way for $O(4)$ regions, simply free the phase of the charge order.

\section*{Phase diagram under magnetic field} 

The measured phase diagram as a function of applied magnetic field ($H$) and temperature ($T$) is particularly remarkable because it shows an abrupt phase transition at $H_{0}=17$ T from a superconducting phase to an incommensurate charge ordered phase \cite{Wu11, LeBoeuf13, Yu12667, Gerber:2015gx, Chang:2016gz}. Both phases have a transition temperature of the same order of magnitude, and co-exist in a small region of the phase diagram. The abruptness of the transition at $H_{0}$ is reminiscent of a spin-flop transition, typical of non-linear $\sigma$-models \cite{Zhang:1997ew,Chakravarty:1988uu}, which could be related to a ``pseudo-spin-flop'' transition of our SU(2) order parameter, sketched in Fig.\ \ref{fig:phase-diagram}.

In the presence of an applied magnetic field, the $O(3)$ non-linear $\sigma$-model can be written as: 
\begin{align}
S  = &\int d^{2}x\left[\frac{\rho_{1}}{2}\left(\partial_{\mu}n_{1}\right)^{2}+\frac{\rho_{\perp}}{2}\sum_{\alpha=2}^3 \left(\overline{\partial}_{\mu}n_{\alpha}\right)^{2}+\frac{H^{2}}{8\pi} \right. \nonumber
\\
& \left. +\sum_{\alpha}m_{\alpha}n_{\alpha}^{2}\right],
\end{align}
where $\overline{\partial}_{\mu}=\partial_{\mu}-qA_{\mu}$, $A_{\mu}$ is the electromagnetic vector potential, $H$ is the applied magnetic field and we have taken $c=1$ and $q=2e/c$. Here again, the vector $n_{\alpha=1,3}$ with $\mathbf{n}^{2}=1$ describes the pseudo-spin states with $n_{1}=\chi$, $n_{2,3}=\Delta,\Delta^{*}$, respectively. $\rho_{1}$ and $\rho_{\perp}$ are the phase stiffnesses which usually are temperature dependent. The masses $m_{\alpha}$ can be taken at zero field such as to favour the superconducting state with for example $m_{1}=0$, and $m_{2,3}<0$.

When $H$ is large enough, vortices with charge modulation core are created until $H_{c2}$ \cite{Wang:2002ke}. Meanwhile, the superconducting state $\Delta$ becomes less favourable compared to the charge modulations $\chi$. An effective homogeneous approximation of the gradient term \cite{Rosenstein2010} yields the spin-flop transition (Fig.\ \ref{fig:phase-diagram}) at a the magnetic field $H_{0} = m_{1}-m_{2,3}$. Interestingly, for fields in the vicinity of the spin-flop transition, the SU(2) symmetry is almost perfectly realised and the model can be mapped onto the attractive Hubbard model at half-filling, known to possess an exact SU(2) particle-hole symmetry \cite{Yang89, Yang:1990cf, Zhang1990}.

A prediction of our model is that inside the charge order phase for $H>H_{c2}$, vortices disappear and dislocations are the only type of topological defects left. The charge order parameter cancels along these dislocations, which means that the superconducting order parameter is then maximal. These ``filaments'' of superconductivity could also be caused by the application of a current \cite{DavisPrivate}, which would be reminiscent of transport measurements where glimpses of superconductivity have been seen at very high fields and low temperatures \cite{Grissonnanche14, Yu12667, Hsu2017}.

\section*{Conclusion}

The complex phenomenology of the cuprates has led to the rise of more and more involved theoretical descriptions, some even untractable analytically. Here, we discussed a simple formalism which naturally gives birth to the wealth of observed phenomena, and enables us to embrace it all at once. We described the SU(2) parameter, which is a non-abelian composite of a charge and a superconducting order parameters, and its derivation from a short-range antiferromagnetic model. This order parameter is constrained by an SU(2) symmetry, which means that it sits on a three-dimensional hypersphere. As the temperature is lowered within the pseudogap phase, some real-space regions lose one charge degree of freedom, and the order parameter thus sits on a two-dimensional sphere. The SU(2) order parameter can then be seen as a pseudo-spin order parameter in these regions. This naturally leads to the formation of pseudo-spin skyrmions, which account for many puzzling features of STM, X-ray, and NMR data. Nematic and time-reversal symmetry-breaking features arise in the other regions where all the degrees of freedom still remain. Finally, we discussed the phase diagram under magnetic field, which could correspond to a pseudo-spin-flop transition.

In summary, general considerations on the SU(2) order parameter allow us to grasp the phase diagram of the cuprate superconductors in its complexity. We think this approach is an important step towards finally unravelling the mysteries of these wonderful materials.

\section*{Acknowledgments}

We thank Y.\ Sidis for stimulating discussions. This work has received
financial support from the ANR project UNESCOS ANR-14-CE05-0007, and
the ERC, under grant agreement AdG-694651-CHAMPAGNE. The authors also
like to thank the IIP (Natal, Brazil), where parts of this work were
done, for hospitality.

\bibliographystyle{apsrev4-1}
\bibliography{Cuprates}

\begin{thebibliography}{83}%
\makeatletter
\providecommand \@ifxundefined [1]{%
 \@ifx{#1\undefined}
}%
\providecommand \@ifnum [1]{%
 \ifnum #1\expandafter \@firstoftwo
 \else \expandafter \@secondoftwo
 \fi
}%
\providecommand \@ifx [1]{%
 \ifx #1\expandafter \@firstoftwo
 \else \expandafter \@secondoftwo
 \fi
}%
\providecommand \natexlab [1]{#1}%
\providecommand \enquote  [1]{``#1''}%
\providecommand \bibnamefont  [1]{#1}%
\providecommand \bibfnamefont [1]{#1}%
\providecommand \citenamefont [1]{#1}%
\providecommand \href@noop [0]{\@secondoftwo}%
\providecommand \href [0]{\begingroup \@sanitize@url \@href}%
\providecommand \@href[1]{\@@startlink{#1}\@@href}%
\providecommand \@@href[1]{\endgroup#1\@@endlink}%
\providecommand \@sanitize@url [0]{\catcode `\\12\catcode `\$12\catcode
  `\&12\catcode `\#12\catcode `\^12\catcode `\_12\catcode `\%12\relax}%
\providecommand \@@startlink[1]{}%
\providecommand \@@endlink[0]{}%
\providecommand \url  [0]{\begingroup\@sanitize@url \@url }%
\providecommand \@url [1]{\endgroup\@href {#1}{\urlprefix }}%
\providecommand \urlprefix  [0]{URL }%
\providecommand \Eprint [0]{\href }%
\providecommand \doibase [0]{http://dx.doi.org/}%
\providecommand \selectlanguage [0]{\@gobble}%
\providecommand \bibinfo  [0]{\@secondoftwo}%
\providecommand \bibfield  [0]{\@secondoftwo}%
\providecommand \translation [1]{[#1]}%
\providecommand \BibitemOpen [0]{}%
\providecommand \bibitemStop [0]{}%
\providecommand \bibitemNoStop [0]{.\EOS\space}%
\providecommand \EOS [0]{\spacefactor3000\relax}%
\providecommand \BibitemShut  [1]{\csname bibitem#1\endcsname}%
\let\auto@bib@innerbib\@empty
%</preamble>
\bibitem [{\citenamefont {Alloul}\ \emph {et~al.}(1989)\citenamefont {Alloul},
  \citenamefont {Ohno},\ and\ \citenamefont {Mendels}}]{Alloul89}%
  \BibitemOpen
  \bibfield  {author} {\bibinfo {author} {\bibfnamefont {H.}~\bibnamefont
  {Alloul}}, \bibinfo {author} {\bibfnamefont {T.}~\bibnamefont {Ohno}}, \ and\
  \bibinfo {author} {\bibfnamefont {P.}~\bibnamefont {Mendels}},\ }\href
  {\doibase 10.1103/PhysRevLett.63.1700} {\bibfield  {journal} {\bibinfo
  {journal} {Phys. Rev. Lett.}\ }\textbf {\bibinfo {volume} {63}},\ \bibinfo
  {pages} {1700} (\bibinfo {year} {1989})}\BibitemShut {NoStop}%
\bibitem [{\citenamefont {Alloul}\ \emph {et~al.}(1991)\citenamefont {Alloul},
  \citenamefont {Mendels}, \citenamefont {Casalta}, \citenamefont {Marucco},\
  and\ \citenamefont {Arabski}}]{Alloul91}%
  \BibitemOpen
  \bibfield  {author} {\bibinfo {author} {\bibfnamefont {H.}~\bibnamefont
  {Alloul}}, \bibinfo {author} {\bibfnamefont {P.}~\bibnamefont {Mendels}},
  \bibinfo {author} {\bibfnamefont {H.}~\bibnamefont {Casalta}}, \bibinfo
  {author} {\bibfnamefont {J.~F.}\ \bibnamefont {Marucco}}, \ and\ \bibinfo
  {author} {\bibfnamefont {J.}~\bibnamefont {Arabski}},\ }\href {\doibase
  10.1103/PhysRevLett.67.3140} {\bibfield  {journal} {\bibinfo  {journal}
  {Phys. Rev. Lett.}\ }\textbf {\bibinfo {volume} {67}},\ \bibinfo {pages}
  {3140} (\bibinfo {year} {1991})}\BibitemShut {NoStop}%
\bibitem [{\citenamefont {Warren}\ \emph {et~al.}(1989)\citenamefont {Warren},
  \citenamefont {Walstedt}, \citenamefont {Brennert}, \citenamefont {Cava},
  \citenamefont {Tycko}, \citenamefont {Bell},\ and\ \citenamefont
  {Dabbagh}}]{Warren89}%
  \BibitemOpen
  \bibfield  {author} {\bibinfo {author} {\bibfnamefont {W.~W.}\ \bibnamefont
  {Warren}}, \bibinfo {author} {\bibfnamefont {R.~E.}\ \bibnamefont
  {Walstedt}}, \bibinfo {author} {\bibfnamefont {G.~F.}\ \bibnamefont
  {Brennert}}, \bibinfo {author} {\bibfnamefont {R.~J.}\ \bibnamefont {Cava}},
  \bibinfo {author} {\bibfnamefont {R.}~\bibnamefont {Tycko}}, \bibinfo
  {author} {\bibfnamefont {R.~F.}\ \bibnamefont {Bell}}, \ and\ \bibinfo
  {author} {\bibfnamefont {G.}~\bibnamefont {Dabbagh}},\ }\href {\doibase
  10.1103/PhysRevLett.62.1193} {\bibfield  {journal} {\bibinfo  {journal}
  {Phys. Rev. Lett.}\ }\textbf {\bibinfo {volume} {62}},\ \bibinfo {pages}
  {1193} (\bibinfo {year} {1989})}\BibitemShut {NoStop}%
\bibitem [{\citenamefont {Campuzano}\ \emph {et~al.}(1998)\citenamefont
  {Campuzano}, \citenamefont {Norman}, \citenamefont {Ding}, \citenamefont
  {Randeria}, \citenamefont {Yokoya}, \citenamefont {Takeuchi}, \citenamefont
  {Takahashi}, \citenamefont {Mochiku}, \citenamefont {Kadowaki}, \citenamefont
  {Guptasarma},\ and\ \citenamefont {Hinks}}]{Campuzano98}%
  \BibitemOpen
  \bibfield  {author} {\bibinfo {author} {\bibfnamefont {J.~C.}\ \bibnamefont
  {Campuzano}}, \bibinfo {author} {\bibfnamefont {M.~R.}\ \bibnamefont
  {Norman}}, \bibinfo {author} {\bibfnamefont {H.}~\bibnamefont {Ding}},
  \bibinfo {author} {\bibfnamefont {M.}~\bibnamefont {Randeria}}, \bibinfo
  {author} {\bibfnamefont {T.}~\bibnamefont {Yokoya}}, \bibinfo {author}
  {\bibfnamefont {T.}~\bibnamefont {Takeuchi}}, \bibinfo {author}
  {\bibfnamefont {T.}~\bibnamefont {Takahashi}}, \bibinfo {author}
  {\bibfnamefont {T.}~\bibnamefont {Mochiku}}, \bibinfo {author} {\bibfnamefont
  {K.}~\bibnamefont {Kadowaki}}, \bibinfo {author} {\bibfnamefont
  {P.}~\bibnamefont {Guptasarma}}, \ and\ \bibinfo {author} {\bibfnamefont
  {D.~G.}\ \bibnamefont {Hinks}},\ }\href {\doibase 10.1038/32366} {\bibfield
  {journal} {\bibinfo  {journal} {Nature}\ }\textbf {\bibinfo {volume} {392}},\
  \bibinfo {pages} {157} (\bibinfo {year} {1998})}\BibitemShut {NoStop}%
\bibitem [{\citenamefont {Vishik}\ \emph {et~al.}(2012)\citenamefont {Vishik},
  \citenamefont {Hashimoto}, \citenamefont {He}, \citenamefont {Lee},
  \citenamefont {Schmitt}, \citenamefont {Lu}, \citenamefont {Moore},
  \citenamefont {Zhang}, \citenamefont {Meevasana}, \citenamefont {Sasagawa},
  \citenamefont {Uchida}, \citenamefont {Fujita}, \citenamefont {Ishida},
  \citenamefont {Ishikado}, \citenamefont {Yoshida}, \citenamefont {Eisaki},
  \citenamefont {Hussain}, \citenamefont {Devereaux},\ and\ \citenamefont
  {Shen}}]{Vishik:2012cc}%
  \BibitemOpen
  \bibfield  {author} {\bibinfo {author} {\bibfnamefont {I.~M.}\ \bibnamefont
  {Vishik}}, \bibinfo {author} {\bibfnamefont {M.}~\bibnamefont {Hashimoto}},
  \bibinfo {author} {\bibfnamefont {R.-H.}\ \bibnamefont {He}}, \bibinfo
  {author} {\bibfnamefont {W.-S.}\ \bibnamefont {Lee}}, \bibinfo {author}
  {\bibfnamefont {F.}~\bibnamefont {Schmitt}}, \bibinfo {author} {\bibfnamefont
  {D.}~\bibnamefont {Lu}}, \bibinfo {author} {\bibfnamefont {R.~G.}\
  \bibnamefont {Moore}}, \bibinfo {author} {\bibfnamefont {C.}~\bibnamefont
  {Zhang}}, \bibinfo {author} {\bibfnamefont {W.}~\bibnamefont {Meevasana}},
  \bibinfo {author} {\bibfnamefont {T.}~\bibnamefont {Sasagawa}}, \bibinfo
  {author} {\bibfnamefont {S.}~\bibnamefont {Uchida}}, \bibinfo {author}
  {\bibfnamefont {K.}~\bibnamefont {Fujita}}, \bibinfo {author} {\bibfnamefont
  {S.}~\bibnamefont {Ishida}}, \bibinfo {author} {\bibfnamefont
  {M.}~\bibnamefont {Ishikado}}, \bibinfo {author} {\bibfnamefont
  {Y.}~\bibnamefont {Yoshida}}, \bibinfo {author} {\bibfnamefont
  {H.}~\bibnamefont {Eisaki}}, \bibinfo {author} {\bibfnamefont
  {Z.}~\bibnamefont {Hussain}}, \bibinfo {author} {\bibfnamefont {T.~P.}\
  \bibnamefont {Devereaux}}, \ and\ \bibinfo {author} {\bibfnamefont {Z.-X.}\
  \bibnamefont {Shen}},\ }\href {\doibase 10.1073/pnas.1209471109} {\bibfield
  {journal} {\bibinfo  {journal} {Proc. Natl. Acad. Sci.}\ }\textbf {\bibinfo
  {volume} {109}},\ \bibinfo {pages} {18332} (\bibinfo {year}
  {2012})}\BibitemShut {NoStop}%
\bibitem [{\citenamefont {Yoshida}\ \emph {et~al.}(2012)\citenamefont
  {Yoshida}, \citenamefont {Hashimoto}, \citenamefont {M~Vishik}, \citenamefont
  {Shen},\ and\ \citenamefont {Fujimori}}]{Yoshida:2012kh}%
  \BibitemOpen
  \bibfield  {author} {\bibinfo {author} {\bibfnamefont {T.}~\bibnamefont
  {Yoshida}}, \bibinfo {author} {\bibfnamefont {M.}~\bibnamefont {Hashimoto}},
  \bibinfo {author} {\bibfnamefont {I.}~\bibnamefont {M~Vishik}}, \bibinfo
  {author} {\bibfnamefont {Z.-X.}\ \bibnamefont {Shen}}, \ and\ \bibinfo
  {author} {\bibfnamefont {A.}~\bibnamefont {Fujimori}},\ }\href {\doibase
  10.1143/JPSJ.81.011006} {\bibfield  {journal} {\bibinfo  {journal} {J. Phys.
  Soc. Jpn.}\ }\textbf {\bibinfo {volume} {81}},\ \bibinfo {pages} {011006}
  (\bibinfo {year} {2012})}\BibitemShut {NoStop}%
\bibitem [{\citenamefont {He}\ \emph {et~al.}(2014)\citenamefont {He},
  \citenamefont {Yin}, \citenamefont {Zech}, \citenamefont {Soumyanarayanan},
  \citenamefont {Yee}, \citenamefont {Williams}, \citenamefont {Boyer},
  \citenamefont {Chatterjee}, \citenamefont {Wise}, \citenamefont {Zeljkovic},
  \citenamefont {Kondo}, \citenamefont {Takeuchi}, \citenamefont {Ikuta},
  \citenamefont {Mistark}, \citenamefont {Markiewicz}, \citenamefont {Bansil},
  \citenamefont {Sachdev}, \citenamefont {Hudson},\ and\ \citenamefont
  {Hoffman}}]{He14}%
  \BibitemOpen
  \bibfield  {author} {\bibinfo {author} {\bibfnamefont {Y.}~\bibnamefont
  {He}}, \bibinfo {author} {\bibfnamefont {Y.}~\bibnamefont {Yin}}, \bibinfo
  {author} {\bibfnamefont {M.}~\bibnamefont {Zech}}, \bibinfo {author}
  {\bibfnamefont {A.}~\bibnamefont {Soumyanarayanan}}, \bibinfo {author}
  {\bibfnamefont {M.~M.}\ \bibnamefont {Yee}}, \bibinfo {author} {\bibfnamefont
  {T.}~\bibnamefont {Williams}}, \bibinfo {author} {\bibfnamefont {M.~C.}\
  \bibnamefont {Boyer}}, \bibinfo {author} {\bibfnamefont {K.}~\bibnamefont
  {Chatterjee}}, \bibinfo {author} {\bibfnamefont {W.~D.}\ \bibnamefont
  {Wise}}, \bibinfo {author} {\bibfnamefont {I.}~\bibnamefont {Zeljkovic}},
  \bibinfo {author} {\bibfnamefont {T.}~\bibnamefont {Kondo}}, \bibinfo
  {author} {\bibfnamefont {T.}~\bibnamefont {Takeuchi}}, \bibinfo {author}
  {\bibfnamefont {H.}~\bibnamefont {Ikuta}}, \bibinfo {author} {\bibfnamefont
  {P.}~\bibnamefont {Mistark}}, \bibinfo {author} {\bibfnamefont {R.~S.}\
  \bibnamefont {Markiewicz}}, \bibinfo {author} {\bibfnamefont
  {A.}~\bibnamefont {Bansil}}, \bibinfo {author} {\bibfnamefont
  {S.}~\bibnamefont {Sachdev}}, \bibinfo {author} {\bibfnamefont {E.~W.}\
  \bibnamefont {Hudson}}, \ and\ \bibinfo {author} {\bibfnamefont {J.~E.}\
  \bibnamefont {Hoffman}},\ }\href {\doibase 10.1126/science.1248221}
  {\bibfield  {journal} {\bibinfo  {journal} {Science}\ }\textbf {\bibinfo
  {volume} {344}},\ \bibinfo {pages} {608} (\bibinfo {year}
  {2014})}\BibitemShut {NoStop}%
\bibitem [{\citenamefont {Vishik}\ \emph {et~al.}(2014)\citenamefont {Vishik},
  \citenamefont {Bari\ifmmode \check{s}\else \v{s}\fi{}i\ifmmode~\acute{c}\else
  \'{c}\fi{}}, \citenamefont {Chan}, \citenamefont {Li}, \citenamefont {Xia},
  \citenamefont {Yu}, \citenamefont {Zhao}, \citenamefont {Lee}, \citenamefont
  {Meevasana}, \citenamefont {Devereaux}, \citenamefont {Greven},\ and\
  \citenamefont {Shen}}]{Vishik14}%
  \BibitemOpen
  \bibfield  {author} {\bibinfo {author} {\bibfnamefont {I.~M.}\ \bibnamefont
  {Vishik}}, \bibinfo {author} {\bibfnamefont {N.}~\bibnamefont {Bari\ifmmode
  \check{s}\else \v{s}\fi{}i\ifmmode~\acute{c}\else \'{c}\fi{}}}, \bibinfo
  {author} {\bibfnamefont {M.~K.}\ \bibnamefont {Chan}}, \bibinfo {author}
  {\bibfnamefont {Y.}~\bibnamefont {Li}}, \bibinfo {author} {\bibfnamefont
  {D.~D.}\ \bibnamefont {Xia}}, \bibinfo {author} {\bibfnamefont
  {G.}~\bibnamefont {Yu}}, \bibinfo {author} {\bibfnamefont {X.}~\bibnamefont
  {Zhao}}, \bibinfo {author} {\bibfnamefont {W.~S.}\ \bibnamefont {Lee}},
  \bibinfo {author} {\bibfnamefont {W.}~\bibnamefont {Meevasana}}, \bibinfo
  {author} {\bibfnamefont {T.~P.}\ \bibnamefont {Devereaux}}, \bibinfo {author}
  {\bibfnamefont {M.}~\bibnamefont {Greven}}, \ and\ \bibinfo {author}
  {\bibfnamefont {Z.-X.}\ \bibnamefont {Shen}},\ }\href {\doibase
  10.1103/PhysRevB.89.195141} {\bibfield  {journal} {\bibinfo  {journal} {Phys.
  Rev. B}\ }\textbf {\bibinfo {volume} {89}},\ \bibinfo {pages} {195141}
  (\bibinfo {year} {2014})}\BibitemShut {NoStop}%
\bibitem [{\citenamefont {Doiron-Leyraud}\ \emph {et~al.}(2007)\citenamefont
  {Doiron-Leyraud}, \citenamefont {Proust}, \citenamefont {LeBoeuf},
  \citenamefont {Levallois}, \citenamefont {Bonnemaison}, \citenamefont
  {Liang}, \citenamefont {Bonn}, \citenamefont {Hardy},\ and\ \citenamefont
  {Taillefer}}]{Doiron-Leyraud07}%
  \BibitemOpen
  \bibfield  {author} {\bibinfo {author} {\bibfnamefont {N.}~\bibnamefont
  {Doiron-Leyraud}}, \bibinfo {author} {\bibfnamefont {C.}~\bibnamefont
  {Proust}}, \bibinfo {author} {\bibfnamefont {D.}~\bibnamefont {LeBoeuf}},
  \bibinfo {author} {\bibfnamefont {J.}~\bibnamefont {Levallois}}, \bibinfo
  {author} {\bibfnamefont {J.-B.}\ \bibnamefont {Bonnemaison}}, \bibinfo
  {author} {\bibfnamefont {R.}~\bibnamefont {Liang}}, \bibinfo {author}
  {\bibfnamefont {D.~A.}\ \bibnamefont {Bonn}}, \bibinfo {author}
  {\bibfnamefont {W.~N.}\ \bibnamefont {Hardy}}, \ and\ \bibinfo {author}
  {\bibfnamefont {L.}~\bibnamefont {Taillefer}},\ }\href {\doibase
  10.1038/nature05872} {\bibfield  {journal} {\bibinfo  {journal} {Nature}\
  }\textbf {\bibinfo {volume} {447}},\ \bibinfo {pages} {565} (\bibinfo {year}
  {2007})}\BibitemShut {NoStop}%
\bibitem [{\citenamefont {LeBoeuf}\ \emph {et~al.}(2007)\citenamefont
  {LeBoeuf}, \citenamefont {Doiron-Leyraud}, \citenamefont {Levallois},
  \citenamefont {Daou}, \citenamefont {Bonnemaison}, \citenamefont {Hussey},
  \citenamefont {Balicas}, \citenamefont {Ramshaw}, \citenamefont {Liang},
  \citenamefont {Bonn}, \citenamefont {Hardy}, \citenamefont {Adachi},
  \citenamefont {Proust},\ and\ \citenamefont {Taillefer}}]{LeBoeuf07}%
  \BibitemOpen
  \bibfield  {author} {\bibinfo {author} {\bibfnamefont {D.}~\bibnamefont
  {LeBoeuf}}, \bibinfo {author} {\bibfnamefont {N.}~\bibnamefont
  {Doiron-Leyraud}}, \bibinfo {author} {\bibfnamefont {J.}~\bibnamefont
  {Levallois}}, \bibinfo {author} {\bibfnamefont {R.}~\bibnamefont {Daou}},
  \bibinfo {author} {\bibfnamefont {J.~B.}\ \bibnamefont {Bonnemaison}},
  \bibinfo {author} {\bibfnamefont {N.~E.}\ \bibnamefont {Hussey}}, \bibinfo
  {author} {\bibfnamefont {L.}~\bibnamefont {Balicas}}, \bibinfo {author}
  {\bibfnamefont {B.~J.}\ \bibnamefont {Ramshaw}}, \bibinfo {author}
  {\bibfnamefont {R.}~\bibnamefont {Liang}}, \bibinfo {author} {\bibfnamefont
  {D.~A.}\ \bibnamefont {Bonn}}, \bibinfo {author} {\bibfnamefont {W.~N.}\
  \bibnamefont {Hardy}}, \bibinfo {author} {\bibfnamefont {S.}~\bibnamefont
  {Adachi}}, \bibinfo {author} {\bibfnamefont {C.}~\bibnamefont {Proust}}, \
  and\ \bibinfo {author} {\bibfnamefont {L.}~\bibnamefont {Taillefer}},\ }\href
  {\doibase 10.1038/nature06332} {\bibfield  {journal} {\bibinfo  {journal}
  {Nature}\ }\textbf {\bibinfo {volume} {450}},\ \bibinfo {pages} {533}
  (\bibinfo {year} {2007})}\BibitemShut {NoStop}%
\bibitem [{\citenamefont {LeBoeuf}\ \emph {et~al.}(2011)\citenamefont
  {LeBoeuf}, \citenamefont {Doiron-Leyraud}, \citenamefont {Vignolle},
  \citenamefont {Sutherland}, \citenamefont {Ramshaw}, \citenamefont
  {Levallois}, \citenamefont {Daou}, \citenamefont {Lalibert\'e}, \citenamefont
  {Cyr-Choini\`ere}, \citenamefont {Chang}, \citenamefont {Jo}, \citenamefont
  {Balicas}, \citenamefont {Liang}, \citenamefont {Bonn}, \citenamefont
  {Hardy}, \citenamefont {Proust},\ and\ \citenamefont
  {Taillefer}}]{LeBoeuf11}%
  \BibitemOpen
  \bibfield  {author} {\bibinfo {author} {\bibfnamefont {D.}~\bibnamefont
  {LeBoeuf}}, \bibinfo {author} {\bibfnamefont {N.}~\bibnamefont
  {Doiron-Leyraud}}, \bibinfo {author} {\bibfnamefont {B.}~\bibnamefont
  {Vignolle}}, \bibinfo {author} {\bibfnamefont {M.}~\bibnamefont
  {Sutherland}}, \bibinfo {author} {\bibfnamefont {B.~J.}\ \bibnamefont
  {Ramshaw}}, \bibinfo {author} {\bibfnamefont {J.}~\bibnamefont {Levallois}},
  \bibinfo {author} {\bibfnamefont {R.}~\bibnamefont {Daou}}, \bibinfo {author}
  {\bibfnamefont {F.}~\bibnamefont {Lalibert\'e}}, \bibinfo {author}
  {\bibfnamefont {O.}~\bibnamefont {Cyr-Choini\`ere}}, \bibinfo {author}
  {\bibfnamefont {J.}~\bibnamefont {Chang}}, \bibinfo {author} {\bibfnamefont
  {Y.~J.}\ \bibnamefont {Jo}}, \bibinfo {author} {\bibfnamefont
  {L.}~\bibnamefont {Balicas}}, \bibinfo {author} {\bibfnamefont
  {R.}~\bibnamefont {Liang}}, \bibinfo {author} {\bibfnamefont {D.~A.}\
  \bibnamefont {Bonn}}, \bibinfo {author} {\bibfnamefont {W.~N.}\ \bibnamefont
  {Hardy}}, \bibinfo {author} {\bibfnamefont {C.}~\bibnamefont {Proust}}, \
  and\ \bibinfo {author} {\bibfnamefont {L.}~\bibnamefont {Taillefer}},\ }\href
  {\doibase 10.1103/PhysRevB.83.054506} {\bibfield  {journal} {\bibinfo
  {journal} {Phys. Rev. B}\ }\textbf {\bibinfo {volume} {83}},\ \bibinfo
  {pages} {054506} (\bibinfo {year} {2011})}\BibitemShut {NoStop}%
\bibitem [{\citenamefont {Lalibert{\'e}}\ \emph {et~al.}(2011)\citenamefont
  {Lalibert{\'e}}, \citenamefont {Chang}, \citenamefont {Doiron-Leyraud},
  \citenamefont {Hassinger}, \citenamefont {Daou}, \citenamefont {Rondeau},
  \citenamefont {Ramshaw}, \citenamefont {Liang}, \citenamefont {Bonn},
  \citenamefont {Hardy}, \citenamefont {Pyon}, \citenamefont {Takayama},
  \citenamefont {Takagi}, \citenamefont {Sheikin}, \citenamefont {Malone},
  \citenamefont {Proust}, \citenamefont {Behnia},\ and\ \citenamefont
  {Taillefer}}]{Laliberte11}%
  \BibitemOpen
  \bibfield  {author} {\bibinfo {author} {\bibfnamefont {F.}~\bibnamefont
  {Lalibert{\'e}}}, \bibinfo {author} {\bibfnamefont {J.}~\bibnamefont
  {Chang}}, \bibinfo {author} {\bibfnamefont {N.}~\bibnamefont
  {Doiron-Leyraud}}, \bibinfo {author} {\bibfnamefont {E.}~\bibnamefont
  {Hassinger}}, \bibinfo {author} {\bibfnamefont {R.}~\bibnamefont {Daou}},
  \bibinfo {author} {\bibfnamefont {M.}~\bibnamefont {Rondeau}}, \bibinfo
  {author} {\bibfnamefont {B.~J.}\ \bibnamefont {Ramshaw}}, \bibinfo {author}
  {\bibfnamefont {R.}~\bibnamefont {Liang}}, \bibinfo {author} {\bibfnamefont
  {D.~A.}\ \bibnamefont {Bonn}}, \bibinfo {author} {\bibfnamefont {W.~N.}\
  \bibnamefont {Hardy}}, \bibinfo {author} {\bibfnamefont {S.}~\bibnamefont
  {Pyon}}, \bibinfo {author} {\bibfnamefont {T.}~\bibnamefont {Takayama}},
  \bibinfo {author} {\bibfnamefont {H.}~\bibnamefont {Takagi}}, \bibinfo
  {author} {\bibfnamefont {I.}~\bibnamefont {Sheikin}}, \bibinfo {author}
  {\bibfnamefont {L.}~\bibnamefont {Malone}}, \bibinfo {author} {\bibfnamefont
  {C.}~\bibnamefont {Proust}}, \bibinfo {author} {\bibfnamefont
  {K.}~\bibnamefont {Behnia}}, \ and\ \bibinfo {author} {\bibfnamefont
  {L.}~\bibnamefont {Taillefer}},\ }\href
  {http://dx.doi.org/10.1038/ncomms1440} {\bibfield  {journal} {\bibinfo
  {journal} {Nat. Commun.}\ }\textbf {\bibinfo {volume} {2}},\ \bibinfo {pages}
  {432} (\bibinfo {year} {2011})}\BibitemShut {NoStop}%
\bibitem [{\citenamefont {Sebastian}\ \emph {et~al.}(2012)\citenamefont
  {Sebastian}, \citenamefont {Harrison}, \citenamefont {Liang}, \citenamefont
  {Bonn}, \citenamefont {Hardy}, \citenamefont {Mielke},\ and\ \citenamefont
  {Lonzarich}}]{Sebastian12}%
  \BibitemOpen
  \bibfield  {author} {\bibinfo {author} {\bibfnamefont {S.~E.}\ \bibnamefont
  {Sebastian}}, \bibinfo {author} {\bibfnamefont {N.}~\bibnamefont {Harrison}},
  \bibinfo {author} {\bibfnamefont {R.}~\bibnamefont {Liang}}, \bibinfo
  {author} {\bibfnamefont {D.~A.}\ \bibnamefont {Bonn}}, \bibinfo {author}
  {\bibfnamefont {W.~N.}\ \bibnamefont {Hardy}}, \bibinfo {author}
  {\bibfnamefont {C.~H.}\ \bibnamefont {Mielke}}, \ and\ \bibinfo {author}
  {\bibfnamefont {G.~G.}\ \bibnamefont {Lonzarich}},\ }\href {\doibase
  10.1103/PhysRevLett.108.196403} {\bibfield  {journal} {\bibinfo  {journal}
  {Phys. Rev. Lett.}\ }\textbf {\bibinfo {volume} {108}},\ \bibinfo {pages}
  {196403} (\bibinfo {year} {2012})}\BibitemShut {NoStop}%
\bibitem [{\citenamefont {Doiron-Leyraud}\ \emph {et~al.}(2013)\citenamefont
  {Doiron-Leyraud}, \citenamefont {Lepault}, \citenamefont {Cyr-Choini\`ere},
  \citenamefont {Vignolle}, \citenamefont {Grissonnanche}, \citenamefont
  {Lalibert\'e}, \citenamefont {Chang}, \citenamefont {Bari\ifmmode
  \check{s}\else \v{s}\fi{}i\ifmmode~\acute{c}\else \'{c}\fi{}}, \citenamefont
  {Chan}, \citenamefont {Ji}, \citenamefont {Zhao}, \citenamefont {Li},
  \citenamefont {Greven}, \citenamefont {Proust},\ and\ \citenamefont
  {Taillefer}}]{Doiron-Leyraud13}%
  \BibitemOpen
  \bibfield  {author} {\bibinfo {author} {\bibfnamefont {N.}~\bibnamefont
  {Doiron-Leyraud}}, \bibinfo {author} {\bibfnamefont {S.}~\bibnamefont
  {Lepault}}, \bibinfo {author} {\bibfnamefont {O.}~\bibnamefont
  {Cyr-Choini\`ere}}, \bibinfo {author} {\bibfnamefont {B.}~\bibnamefont
  {Vignolle}}, \bibinfo {author} {\bibfnamefont {G.}~\bibnamefont
  {Grissonnanche}}, \bibinfo {author} {\bibfnamefont {F.}~\bibnamefont
  {Lalibert\'e}}, \bibinfo {author} {\bibfnamefont {J.}~\bibnamefont {Chang}},
  \bibinfo {author} {\bibfnamefont {N.}~\bibnamefont {Bari\ifmmode
  \check{s}\else \v{s}\fi{}i\ifmmode~\acute{c}\else \'{c}\fi{}}}, \bibinfo
  {author} {\bibfnamefont {M.~K.}\ \bibnamefont {Chan}}, \bibinfo {author}
  {\bibfnamefont {L.}~\bibnamefont {Ji}}, \bibinfo {author} {\bibfnamefont
  {X.}~\bibnamefont {Zhao}}, \bibinfo {author} {\bibfnamefont {Y.}~\bibnamefont
  {Li}}, \bibinfo {author} {\bibfnamefont {M.}~\bibnamefont {Greven}}, \bibinfo
  {author} {\bibfnamefont {C.}~\bibnamefont {Proust}}, \ and\ \bibinfo {author}
  {\bibfnamefont {L.}~\bibnamefont {Taillefer}},\ }\href {\doibase
  10.1103/PhysRevX.3.021019} {\bibfield  {journal} {\bibinfo  {journal} {Phys.
  Rev. X}\ }\textbf {\bibinfo {volume} {3}},\ \bibinfo {pages} {021019}
  (\bibinfo {year} {2013})}\BibitemShut {NoStop}%
\bibitem [{\citenamefont {{Bari{\v s}i{\'c}}}\ \emph
  {et~al.}(2013)\citenamefont {{Bari{\v s}i{\'c}}}, \citenamefont {{Badoux}},
  \citenamefont {{Chan}}, \citenamefont {{Dorow}}, \citenamefont {{Tabis}},
  \citenamefont {{Vignolle}}, \citenamefont {{Yu}}, \citenamefont
  {{B{\'e}ard}}, \citenamefont {{Zhao}}, \citenamefont {{Proust}},\ and\
  \citenamefont {{Greven}}}]{Barisic2013}%
  \BibitemOpen
  \bibfield  {author} {\bibinfo {author} {\bibfnamefont {N.}~\bibnamefont
  {{Bari{\v s}i{\'c}}}}, \bibinfo {author} {\bibfnamefont {S.}~\bibnamefont
  {{Badoux}}}, \bibinfo {author} {\bibfnamefont {M.~K.}\ \bibnamefont
  {{Chan}}}, \bibinfo {author} {\bibfnamefont {C.}~\bibnamefont {{Dorow}}},
  \bibinfo {author} {\bibfnamefont {W.}~\bibnamefont {{Tabis}}}, \bibinfo
  {author} {\bibfnamefont {B.}~\bibnamefont {{Vignolle}}}, \bibinfo {author}
  {\bibfnamefont {G.}~\bibnamefont {{Yu}}}, \bibinfo {author} {\bibfnamefont
  {J.}~\bibnamefont {{B{\'e}ard}}}, \bibinfo {author} {\bibfnamefont
  {X.}~\bibnamefont {{Zhao}}}, \bibinfo {author} {\bibfnamefont
  {C.}~\bibnamefont {{Proust}}}, \ and\ \bibinfo {author} {\bibfnamefont
  {M.}~\bibnamefont {{Greven}}},\ }\href {\doibase 10.1038/nphys2792}
  {\bibfield  {journal} {\bibinfo  {journal} {Nature Physics}\ }\textbf
  {\bibinfo {volume} {9}},\ \bibinfo {pages} {761} (\bibinfo {year}
  {2013})}\BibitemShut {NoStop}%
\bibitem [{\citenamefont {{Grissonnanche}}\ \emph {et~al.}(2015)\citenamefont
  {{Grissonnanche}}, \citenamefont {{Laliberte}}, \citenamefont
  {{Dufour-Beausejour}}, \citenamefont {{Riopel}}, \citenamefont {{Badoux}},
  \citenamefont {{Caouette-Mansour}}, \citenamefont {{Matusiak}}, \citenamefont
  {{Juneau-Fecteau}}, \citenamefont {{Bourgeois-Hope}}, \citenamefont
  {{Cyr-Choiniere}}, \citenamefont {{Baglo}}, \citenamefont {{Ramshaw}},
  \citenamefont {{Liang}}, \citenamefont {{Bonn}}, \citenamefont {{Hardy}},
  \citenamefont {{Kramer}}, \citenamefont {{LeBoeuf}}, \citenamefont {{Graf}},
  \citenamefont {{Doiron-Leyraud}},\ and\ \citenamefont
  {{Taillefer}}}]{Grissonnanche:2015tl}%
  \BibitemOpen
  \bibfield  {author} {\bibinfo {author} {\bibfnamefont {G.}~\bibnamefont
  {{Grissonnanche}}}, \bibinfo {author} {\bibfnamefont {F.}~\bibnamefont
  {{Laliberte}}}, \bibinfo {author} {\bibfnamefont {S.}~\bibnamefont
  {{Dufour-Beausejour}}}, \bibinfo {author} {\bibfnamefont {A.}~\bibnamefont
  {{Riopel}}}, \bibinfo {author} {\bibfnamefont {S.}~\bibnamefont {{Badoux}}},
  \bibinfo {author} {\bibfnamefont {M.}~\bibnamefont {{Caouette-Mansour}}},
  \bibinfo {author} {\bibfnamefont {M.}~\bibnamefont {{Matusiak}}}, \bibinfo
  {author} {\bibfnamefont {A.}~\bibnamefont {{Juneau-Fecteau}}}, \bibinfo
  {author} {\bibfnamefont {P.}~\bibnamefont {{Bourgeois-Hope}}}, \bibinfo
  {author} {\bibfnamefont {O.}~\bibnamefont {{Cyr-Choiniere}}}, \bibinfo
  {author} {\bibfnamefont {J.~C.}\ \bibnamefont {{Baglo}}}, \bibinfo {author}
  {\bibfnamefont {B.~J.}\ \bibnamefont {{Ramshaw}}}, \bibinfo {author}
  {\bibfnamefont {R.}~\bibnamefont {{Liang}}}, \bibinfo {author} {\bibfnamefont
  {D.~A.}\ \bibnamefont {{Bonn}}}, \bibinfo {author} {\bibfnamefont {W.~N.}\
  \bibnamefont {{Hardy}}}, \bibinfo {author} {\bibfnamefont {S.}~\bibnamefont
  {{Kramer}}}, \bibinfo {author} {\bibfnamefont {D.}~\bibnamefont {{LeBoeuf}}},
  \bibinfo {author} {\bibfnamefont {D.}~\bibnamefont {{Graf}}}, \bibinfo
  {author} {\bibfnamefont {N.}~\bibnamefont {{Doiron-Leyraud}}}, \ and\
  \bibinfo {author} {\bibfnamefont {L.}~\bibnamefont {{Taillefer}}},\ }\href
  {http://arxiv.org/abs/1508.05486} {\  (\bibinfo {year} {2015})},\ \Eprint
  {http://arxiv.org/abs/1508.05486} {arXiv:1508.05486} \BibitemShut {NoStop}%
\bibitem [{\citenamefont {Ghiringhelli}\ \emph {et~al.}(2012)\citenamefont
  {Ghiringhelli}, \citenamefont {Le~Tacon}, \citenamefont {Minola},
  \citenamefont {Blanco-Canosa}, \citenamefont {Mazzoli}, \citenamefont
  {Brookes}, \citenamefont {De~Luca}, \citenamefont {Frano}, \citenamefont
  {Hawthorn}, \citenamefont {He}, \citenamefont {Loew}, \citenamefont {Sala},
  \citenamefont {Peets}, \citenamefont {Salluzzo}, \citenamefont {Schierle},
  \citenamefont {Sutarto}, \citenamefont {Sawatzky}, \citenamefont {Weschke},
  \citenamefont {Keimer},\ and\ \citenamefont {Braicovich}}]{Ghiringhelli12}%
  \BibitemOpen
  \bibfield  {author} {\bibinfo {author} {\bibfnamefont {G.}~\bibnamefont
  {Ghiringhelli}}, \bibinfo {author} {\bibfnamefont {M.}~\bibnamefont
  {Le~Tacon}}, \bibinfo {author} {\bibfnamefont {M.}~\bibnamefont {Minola}},
  \bibinfo {author} {\bibfnamefont {S.}~\bibnamefont {Blanco-Canosa}}, \bibinfo
  {author} {\bibfnamefont {C.}~\bibnamefont {Mazzoli}}, \bibinfo {author}
  {\bibfnamefont {N.~B.}\ \bibnamefont {Brookes}}, \bibinfo {author}
  {\bibfnamefont {G.~M.}\ \bibnamefont {De~Luca}}, \bibinfo {author}
  {\bibfnamefont {A.}~\bibnamefont {Frano}}, \bibinfo {author} {\bibfnamefont
  {D.~G.}\ \bibnamefont {Hawthorn}}, \bibinfo {author} {\bibfnamefont
  {F.}~\bibnamefont {He}}, \bibinfo {author} {\bibfnamefont {T.}~\bibnamefont
  {Loew}}, \bibinfo {author} {\bibfnamefont {M.~M.}\ \bibnamefont {Sala}},
  \bibinfo {author} {\bibfnamefont {D.~C.}\ \bibnamefont {Peets}}, \bibinfo
  {author} {\bibfnamefont {M.}~\bibnamefont {Salluzzo}}, \bibinfo {author}
  {\bibfnamefont {E.}~\bibnamefont {Schierle}}, \bibinfo {author}
  {\bibfnamefont {R.}~\bibnamefont {Sutarto}}, \bibinfo {author} {\bibfnamefont
  {G.~A.}\ \bibnamefont {Sawatzky}}, \bibinfo {author} {\bibfnamefont
  {E.}~\bibnamefont {Weschke}}, \bibinfo {author} {\bibfnamefont
  {B.}~\bibnamefont {Keimer}}, \ and\ \bibinfo {author} {\bibfnamefont
  {L.}~\bibnamefont {Braicovich}},\ }\href {\doibase 10.1126/science.1223532}
  {\bibfield  {journal} {\bibinfo  {journal} {Science}\ }\textbf {\bibinfo
  {volume} {337}},\ \bibinfo {pages} {821} (\bibinfo {year}
  {2012})}\BibitemShut {NoStop}%
\bibitem [{\citenamefont {Chang}\ \emph {et~al.}(2012)\citenamefont {Chang},
  \citenamefont {Blackburn}, \citenamefont {Holmes}, \citenamefont
  {Christensen}, \citenamefont {Larsen}, \citenamefont {Mesot}, \citenamefont
  {Liang}, \citenamefont {Bonn}, \citenamefont {Hardy}, \citenamefont
  {Watenphul}, \citenamefont {Zimmermann}, \citenamefont {Forgan},\ and\
  \citenamefont {Hayden}}]{Chang12}%
  \BibitemOpen
  \bibfield  {author} {\bibinfo {author} {\bibfnamefont {J.}~\bibnamefont
  {Chang}}, \bibinfo {author} {\bibfnamefont {E.}~\bibnamefont {Blackburn}},
  \bibinfo {author} {\bibfnamefont {A.~T.}\ \bibnamefont {Holmes}}, \bibinfo
  {author} {\bibfnamefont {N.~B.}\ \bibnamefont {Christensen}}, \bibinfo
  {author} {\bibfnamefont {J.}~\bibnamefont {Larsen}}, \bibinfo {author}
  {\bibfnamefont {J.}~\bibnamefont {Mesot}}, \bibinfo {author} {\bibfnamefont
  {R.}~\bibnamefont {Liang}}, \bibinfo {author} {\bibfnamefont {D.~A.}\
  \bibnamefont {Bonn}}, \bibinfo {author} {\bibfnamefont {W.~N.}\ \bibnamefont
  {Hardy}}, \bibinfo {author} {\bibfnamefont {A.}~\bibnamefont {Watenphul}},
  \bibinfo {author} {\bibfnamefont {M.~v.}\ \bibnamefont {Zimmermann}},
  \bibinfo {author} {\bibfnamefont {E.~M.}\ \bibnamefont {Forgan}}, \ and\
  \bibinfo {author} {\bibfnamefont {S.~M.}\ \bibnamefont {Hayden}},\ }\href
  {\doibase 10.1038/nphys2456} {\bibfield  {journal} {\bibinfo  {journal} {Nat
  Phys}\ }\textbf {\bibinfo {volume} {8}},\ \bibinfo {pages} {871} (\bibinfo
  {year} {2012})}\BibitemShut {NoStop}%
\bibitem [{\citenamefont {Achkar}\ \emph {et~al.}(2012)\citenamefont {Achkar},
  \citenamefont {Sutarto}, \citenamefont {Mao}, \citenamefont {He},
  \citenamefont {Frano}, \citenamefont {Blanco-Canosa}, \citenamefont
  {Le~Tacon}, \citenamefont {Ghiringhelli}, \citenamefont {Braicovich},
  \citenamefont {Minola}, \citenamefont {Moretti~Sala}, \citenamefont
  {Mazzoli}, \citenamefont {Liang}, \citenamefont {Bonn}, \citenamefont
  {Hardy}, \citenamefont {Keimer}, \citenamefont {Sawatzky},\ and\
  \citenamefont {Hawthorn}}]{Achkar12}%
  \BibitemOpen
  \bibfield  {author} {\bibinfo {author} {\bibfnamefont {A.~J.}\ \bibnamefont
  {Achkar}}, \bibinfo {author} {\bibfnamefont {R.}~\bibnamefont {Sutarto}},
  \bibinfo {author} {\bibfnamefont {X.}~\bibnamefont {Mao}}, \bibinfo {author}
  {\bibfnamefont {F.}~\bibnamefont {He}}, \bibinfo {author} {\bibfnamefont
  {A.}~\bibnamefont {Frano}}, \bibinfo {author} {\bibfnamefont
  {S.}~\bibnamefont {Blanco-Canosa}}, \bibinfo {author} {\bibfnamefont
  {M.}~\bibnamefont {Le~Tacon}}, \bibinfo {author} {\bibfnamefont
  {G.}~\bibnamefont {Ghiringhelli}}, \bibinfo {author} {\bibfnamefont
  {L.}~\bibnamefont {Braicovich}}, \bibinfo {author} {\bibfnamefont
  {M.}~\bibnamefont {Minola}}, \bibinfo {author} {\bibfnamefont
  {M.}~\bibnamefont {Moretti~Sala}}, \bibinfo {author} {\bibfnamefont
  {C.}~\bibnamefont {Mazzoli}}, \bibinfo {author} {\bibfnamefont
  {R.}~\bibnamefont {Liang}}, \bibinfo {author} {\bibfnamefont {D.~A.}\
  \bibnamefont {Bonn}}, \bibinfo {author} {\bibfnamefont {W.~N.}\ \bibnamefont
  {Hardy}}, \bibinfo {author} {\bibfnamefont {B.}~\bibnamefont {Keimer}},
  \bibinfo {author} {\bibfnamefont {G.~A.}\ \bibnamefont {Sawatzky}}, \ and\
  \bibinfo {author} {\bibfnamefont {D.~G.}\ \bibnamefont {Hawthorn}},\ }\href
  {\doibase 10.1103/PhysRevLett.109.167001} {\bibfield  {journal} {\bibinfo
  {journal} {Phys. Rev. Lett.}\ }\textbf {\bibinfo {volume} {109}},\ \bibinfo
  {pages} {167001} (\bibinfo {year} {2012})}\BibitemShut {NoStop}%
\bibitem [{\citenamefont {Blanco-Canosa}\ \emph {et~al.}(2013)\citenamefont
  {Blanco-Canosa}, \citenamefont {Frano}, \citenamefont {Loew}, \citenamefont
  {Lu}, \citenamefont {Porras}, \citenamefont {Ghiringhelli}, \citenamefont
  {Minola}, \citenamefont {Mazzoli}, \citenamefont {Braicovich}, \citenamefont
  {Schierle}, \citenamefont {Weschke}, \citenamefont {Le~Tacon},\ and\
  \citenamefont {Keimer}}]{Blanco-Canosa13}%
  \BibitemOpen
  \bibfield  {author} {\bibinfo {author} {\bibfnamefont {S.}~\bibnamefont
  {Blanco-Canosa}}, \bibinfo {author} {\bibfnamefont {A.}~\bibnamefont
  {Frano}}, \bibinfo {author} {\bibfnamefont {T.}~\bibnamefont {Loew}},
  \bibinfo {author} {\bibfnamefont {Y.}~\bibnamefont {Lu}}, \bibinfo {author}
  {\bibfnamefont {J.}~\bibnamefont {Porras}}, \bibinfo {author} {\bibfnamefont
  {G.}~\bibnamefont {Ghiringhelli}}, \bibinfo {author} {\bibfnamefont
  {M.}~\bibnamefont {Minola}}, \bibinfo {author} {\bibfnamefont
  {C.}~\bibnamefont {Mazzoli}}, \bibinfo {author} {\bibfnamefont
  {L.}~\bibnamefont {Braicovich}}, \bibinfo {author} {\bibfnamefont
  {E.}~\bibnamefont {Schierle}}, \bibinfo {author} {\bibfnamefont
  {E.}~\bibnamefont {Weschke}}, \bibinfo {author} {\bibfnamefont
  {M.}~\bibnamefont {Le~Tacon}}, \ and\ \bibinfo {author} {\bibfnamefont
  {B.}~\bibnamefont {Keimer}},\ }\href {\doibase
  10.1103/PhysRevLett.110.187001} {\bibfield  {journal} {\bibinfo  {journal}
  {Phys. Rev. Lett.}\ }\textbf {\bibinfo {volume} {110}},\ \bibinfo {pages}
  {187001} (\bibinfo {year} {2013})}\BibitemShut {NoStop}%
\bibitem [{\citenamefont {Blackburn}\ \emph
  {et~al.}(2013{\natexlab{a}})\citenamefont {Blackburn}, \citenamefont {Chang},
  \citenamefont {H\"ucker}, \citenamefont {Holmes}, \citenamefont
  {Christensen}, \citenamefont {Liang}, \citenamefont {Bonn}, \citenamefont
  {Hardy}, \citenamefont {R\"utt}, \citenamefont {Gutowski}, \citenamefont
  {Zimmermann}, \citenamefont {Forgan},\ and\ \citenamefont
  {Hayden}}]{Blackburn13a}%
  \BibitemOpen
  \bibfield  {author} {\bibinfo {author} {\bibfnamefont {E.}~\bibnamefont
  {Blackburn}}, \bibinfo {author} {\bibfnamefont {J.}~\bibnamefont {Chang}},
  \bibinfo {author} {\bibfnamefont {M.}~\bibnamefont {H\"ucker}}, \bibinfo
  {author} {\bibfnamefont {A.~T.}\ \bibnamefont {Holmes}}, \bibinfo {author}
  {\bibfnamefont {N.~B.}\ \bibnamefont {Christensen}}, \bibinfo {author}
  {\bibfnamefont {R.}~\bibnamefont {Liang}}, \bibinfo {author} {\bibfnamefont
  {D.~A.}\ \bibnamefont {Bonn}}, \bibinfo {author} {\bibfnamefont {W.~N.}\
  \bibnamefont {Hardy}}, \bibinfo {author} {\bibfnamefont {U.}~\bibnamefont
  {R\"utt}}, \bibinfo {author} {\bibfnamefont {O.}~\bibnamefont {Gutowski}},
  \bibinfo {author} {\bibfnamefont {M.~v.}\ \bibnamefont {Zimmermann}},
  \bibinfo {author} {\bibfnamefont {E.~M.}\ \bibnamefont {Forgan}}, \ and\
  \bibinfo {author} {\bibfnamefont {S.~M.}\ \bibnamefont {Hayden}},\ }\href
  {\doibase 10.1103/PhysRevLett.110.137004} {\bibfield  {journal} {\bibinfo
  {journal} {Phys. Rev. Lett.}\ }\textbf {\bibinfo {volume} {110}},\ \bibinfo
  {pages} {137004} (\bibinfo {year} {2013}{\natexlab{a}})}\BibitemShut
  {NoStop}%
\bibitem [{\citenamefont {Blackburn}\ \emph
  {et~al.}(2013{\natexlab{b}})\citenamefont {Blackburn}, \citenamefont {Chang},
  \citenamefont {Said}, \citenamefont {Leu}, \citenamefont {Liang},
  \citenamefont {Bonn}, \citenamefont {Hardy}, \citenamefont {Forgan},\ and\
  \citenamefont {Hayden}}]{Blackburn13b}%
  \BibitemOpen
  \bibfield  {author} {\bibinfo {author} {\bibfnamefont {E.}~\bibnamefont
  {Blackburn}}, \bibinfo {author} {\bibfnamefont {J.}~\bibnamefont {Chang}},
  \bibinfo {author} {\bibfnamefont {A.~H.}\ \bibnamefont {Said}}, \bibinfo
  {author} {\bibfnamefont {B.~M.}\ \bibnamefont {Leu}}, \bibinfo {author}
  {\bibfnamefont {R.}~\bibnamefont {Liang}}, \bibinfo {author} {\bibfnamefont
  {D.~A.}\ \bibnamefont {Bonn}}, \bibinfo {author} {\bibfnamefont {W.~N.}\
  \bibnamefont {Hardy}}, \bibinfo {author} {\bibfnamefont {E.~M.}\ \bibnamefont
  {Forgan}}, \ and\ \bibinfo {author} {\bibfnamefont {S.~M.}\ \bibnamefont
  {Hayden}},\ }\href {\doibase 10.1103/PhysRevB.88.054506} {\bibfield
  {journal} {\bibinfo  {journal} {Phys. Rev. B}\ }\textbf {\bibinfo {volume}
  {88}},\ \bibinfo {pages} {054506} (\bibinfo {year}
  {2013}{\natexlab{b}})}\BibitemShut {NoStop}%
\bibitem [{\citenamefont {Thampy}\ \emph {et~al.}(2013)\citenamefont {Thampy},
  \citenamefont {Blanco-Canosa}, \citenamefont {Garc\'{i}a-Fern\'andez},
  \citenamefont {Dean}, \citenamefont {Gu}, \citenamefont {F\"orst},
  \citenamefont {Loew}, \citenamefont {Keimer}, \citenamefont {Le~Tacon},
  \citenamefont {Wilkins},\ and\ \citenamefont {Hill}}]{Thampy13}%
  \BibitemOpen
  \bibfield  {author} {\bibinfo {author} {\bibfnamefont {V.}~\bibnamefont
  {Thampy}}, \bibinfo {author} {\bibfnamefont {S.}~\bibnamefont
  {Blanco-Canosa}}, \bibinfo {author} {\bibfnamefont {M.}~\bibnamefont
  {Garc\'{i}a-Fern\'andez}}, \bibinfo {author} {\bibfnamefont {M.~P.~M.}\
  \bibnamefont {Dean}}, \bibinfo {author} {\bibfnamefont {G.~D.}\ \bibnamefont
  {Gu}}, \bibinfo {author} {\bibfnamefont {M.}~\bibnamefont {F\"orst}},
  \bibinfo {author} {\bibfnamefont {T.}~\bibnamefont {Loew}}, \bibinfo {author}
  {\bibfnamefont {B.}~\bibnamefont {Keimer}}, \bibinfo {author} {\bibfnamefont
  {M.}~\bibnamefont {Le~Tacon}}, \bibinfo {author} {\bibfnamefont {S.~B.}\
  \bibnamefont {Wilkins}}, \ and\ \bibinfo {author} {\bibfnamefont {J.~P.}\
  \bibnamefont {Hill}},\ }\href {\doibase 10.1103/PhysRevB.88.024505}
  {\bibfield  {journal} {\bibinfo  {journal} {Phys. Rev. B}\ }\textbf {\bibinfo
  {volume} {88}},\ \bibinfo {pages} {024505} (\bibinfo {year}
  {2013})}\BibitemShut {NoStop}%
\bibitem [{\citenamefont {Blanco-Canosa}\ \emph {et~al.}(2014)\citenamefont
  {Blanco-Canosa}, \citenamefont {Frano}, \citenamefont {Schierle},
  \citenamefont {Porras}, \citenamefont {Loew}, \citenamefont {Minola},
  \citenamefont {Bluschke}, \citenamefont {Weschke}, \citenamefont {Keimer},\
  and\ \citenamefont {Le~Tacon}}]{Blanco-Canosa14}%
  \BibitemOpen
  \bibfield  {author} {\bibinfo {author} {\bibfnamefont {S.}~\bibnamefont
  {Blanco-Canosa}}, \bibinfo {author} {\bibfnamefont {A.}~\bibnamefont
  {Frano}}, \bibinfo {author} {\bibfnamefont {E.}~\bibnamefont {Schierle}},
  \bibinfo {author} {\bibfnamefont {J.}~\bibnamefont {Porras}}, \bibinfo
  {author} {\bibfnamefont {T.}~\bibnamefont {Loew}}, \bibinfo {author}
  {\bibfnamefont {M.}~\bibnamefont {Minola}}, \bibinfo {author} {\bibfnamefont
  {M.}~\bibnamefont {Bluschke}}, \bibinfo {author} {\bibfnamefont
  {E.}~\bibnamefont {Weschke}}, \bibinfo {author} {\bibfnamefont
  {B.}~\bibnamefont {Keimer}}, \ and\ \bibinfo {author} {\bibfnamefont
  {M.}~\bibnamefont {Le~Tacon}},\ }\href {\doibase 10.1103/PhysRevB.90.054513}
  {\bibfield  {journal} {\bibinfo  {journal} {Phys. Rev. B}\ }\textbf {\bibinfo
  {volume} {90}},\ \bibinfo {pages} {054513} (\bibinfo {year}
  {2014})}\BibitemShut {NoStop}%
\bibitem [{\citenamefont {Tabis}\ \emph {et~al.}(2014)\citenamefont {Tabis},
  \citenamefont {Li}, \citenamefont {{Le Tacon}}, \citenamefont {Braicovich},
  \citenamefont {Kreyssig}, \citenamefont {Minola}, \citenamefont {Dellea},
  \citenamefont {Weschke}, \citenamefont {Veit}, \citenamefont {Ramazanoglu},
  \citenamefont {Goldman}, \citenamefont {Schmitt}, \citenamefont
  {Ghiringhelli}, \citenamefont {{Bari{\v s}i{\'c}}}, \citenamefont {Chan},
  \citenamefont {Dorow}, \citenamefont {Yu}, \citenamefont {Zhao},
  \citenamefont {Keimer},\ and\ \citenamefont {Greven}}]{Tabis14}%
  \BibitemOpen
  \bibfield  {author} {\bibinfo {author} {\bibfnamefont {W.}~\bibnamefont
  {Tabis}}, \bibinfo {author} {\bibfnamefont {Y.}~\bibnamefont {Li}}, \bibinfo
  {author} {\bibfnamefont {M.}~\bibnamefont {{Le Tacon}}}, \bibinfo {author}
  {\bibfnamefont {L.}~\bibnamefont {Braicovich}}, \bibinfo {author}
  {\bibfnamefont {A.}~\bibnamefont {Kreyssig}}, \bibinfo {author}
  {\bibfnamefont {M.}~\bibnamefont {Minola}}, \bibinfo {author} {\bibfnamefont
  {G.}~\bibnamefont {Dellea}}, \bibinfo {author} {\bibfnamefont
  {E.}~\bibnamefont {Weschke}}, \bibinfo {author} {\bibfnamefont {M.~J.}\
  \bibnamefont {Veit}}, \bibinfo {author} {\bibfnamefont {M.}~\bibnamefont
  {Ramazanoglu}}, \bibinfo {author} {\bibfnamefont {A.~I.}\ \bibnamefont
  {Goldman}}, \bibinfo {author} {\bibfnamefont {T.}~\bibnamefont {Schmitt}},
  \bibinfo {author} {\bibfnamefont {G.}~\bibnamefont {Ghiringhelli}}, \bibinfo
  {author} {\bibfnamefont {N.}~\bibnamefont {{Bari{\v s}i{\'c}}}}, \bibinfo
  {author} {\bibfnamefont {M.~K.}\ \bibnamefont {Chan}}, \bibinfo {author}
  {\bibfnamefont {C.~J.}\ \bibnamefont {Dorow}}, \bibinfo {author}
  {\bibfnamefont {G.}~\bibnamefont {Yu}}, \bibinfo {author} {\bibfnamefont
  {X.}~\bibnamefont {Zhao}}, \bibinfo {author} {\bibfnamefont {B.}~\bibnamefont
  {Keimer}}, \ and\ \bibinfo {author} {\bibfnamefont {M.}~\bibnamefont
  {Greven}},\ }\href {\doibase 10.1038/ncomms6875} {\bibfield  {journal}
  {\bibinfo  {journal} {Nat. Commun.}\ }\textbf {\bibinfo {volume} {5}},\
  \bibinfo {pages} {5875} (\bibinfo {year} {2014})}\BibitemShut {NoStop}%
\bibitem [{\citenamefont {Comin}\ \emph {et~al.}(2014)\citenamefont {Comin},
  \citenamefont {Frano}, \citenamefont {Yee}, \citenamefont {Yoshida},
  \citenamefont {Eisaki}, \citenamefont {Schierle}, \citenamefont {Weschke},
  \citenamefont {Sutarto}, \citenamefont {He}, \citenamefont {Soumyanarayanan},
  \citenamefont {He}, \citenamefont {Le~Tacon}, \citenamefont {Elfimov},
  \citenamefont {Hoffman}, \citenamefont {Sawatzky}, \citenamefont {Keimer},\
  and\ \citenamefont {Damascelli}}]{Comin14}%
  \BibitemOpen
  \bibfield  {author} {\bibinfo {author} {\bibfnamefont {R.}~\bibnamefont
  {Comin}}, \bibinfo {author} {\bibfnamefont {A.}~\bibnamefont {Frano}},
  \bibinfo {author} {\bibfnamefont {M.~M.}\ \bibnamefont {Yee}}, \bibinfo
  {author} {\bibfnamefont {Y.}~\bibnamefont {Yoshida}}, \bibinfo {author}
  {\bibfnamefont {H.}~\bibnamefont {Eisaki}}, \bibinfo {author} {\bibfnamefont
  {E.}~\bibnamefont {Schierle}}, \bibinfo {author} {\bibfnamefont
  {E.}~\bibnamefont {Weschke}}, \bibinfo {author} {\bibfnamefont
  {R.}~\bibnamefont {Sutarto}}, \bibinfo {author} {\bibfnamefont
  {F.}~\bibnamefont {He}}, \bibinfo {author} {\bibfnamefont {A.}~\bibnamefont
  {Soumyanarayanan}}, \bibinfo {author} {\bibfnamefont {Y.}~\bibnamefont {He}},
  \bibinfo {author} {\bibfnamefont {M.}~\bibnamefont {Le~Tacon}}, \bibinfo
  {author} {\bibfnamefont {I.~S.}\ \bibnamefont {Elfimov}}, \bibinfo {author}
  {\bibfnamefont {J.~E.}\ \bibnamefont {Hoffman}}, \bibinfo {author}
  {\bibfnamefont {G.~A.}\ \bibnamefont {Sawatzky}}, \bibinfo {author}
  {\bibfnamefont {B.}~\bibnamefont {Keimer}}, \ and\ \bibinfo {author}
  {\bibfnamefont {A.}~\bibnamefont {Damascelli}},\ }\href {\doibase
  10.1126/science.1242996} {\bibfield  {journal} {\bibinfo  {journal}
  {Science}\ }\textbf {\bibinfo {volume} {343}},\ \bibinfo {pages} {390}
  (\bibinfo {year} {2014})}\BibitemShut {NoStop}%
\bibitem [{\citenamefont {Comin}\ and\ \citenamefont
  {Damascelli}(2016)}]{Comin:2015vc}%
  \BibitemOpen
  \bibfield  {author} {\bibinfo {author} {\bibfnamefont {R.}~\bibnamefont
  {Comin}}\ and\ \bibinfo {author} {\bibfnamefont {A.}~\bibnamefont
  {Damascelli}},\ }\href {\doibase 10.1146/annurev-conmatphys-031115-011401}
  {\bibfield  {journal} {\bibinfo  {journal} {Annual Review of Condensed Matter
  Physics}\ }\textbf {\bibinfo {volume} {7}},\ \bibinfo {pages} {369} (\bibinfo
  {year} {2016})}\BibitemShut {NoStop}%
\bibitem [{\citenamefont {Comin}\ \emph {et~al.}(2015)\citenamefont {Comin},
  \citenamefont {Sutarto}, \citenamefont {He}, \citenamefont {da~Silva~Neto},
  \citenamefont {Chauviere}, \citenamefont {Frano}, \citenamefont {Liang},
  \citenamefont {Hardy}, \citenamefont {Bonn}, \citenamefont {Yoshida},
  \citenamefont {Eisaki}, \citenamefont {Achkar}, \citenamefont {Hawthorn},
  \citenamefont {Keimer}, \citenamefont {Sawatzky},\ and\ \citenamefont
  {Damascelli}}]{Comin:2015ca}%
  \BibitemOpen
  \bibfield  {author} {\bibinfo {author} {\bibfnamefont {R.}~\bibnamefont
  {Comin}}, \bibinfo {author} {\bibfnamefont {R.}~\bibnamefont {Sutarto}},
  \bibinfo {author} {\bibfnamefont {F.}~\bibnamefont {He}}, \bibinfo {author}
  {\bibfnamefont {E.~H.}\ \bibnamefont {da~Silva~Neto}}, \bibinfo {author}
  {\bibfnamefont {L.}~\bibnamefont {Chauviere}}, \bibinfo {author}
  {\bibfnamefont {A.}~\bibnamefont {Frano}}, \bibinfo {author} {\bibfnamefont
  {R.}~\bibnamefont {Liang}}, \bibinfo {author} {\bibfnamefont {W.~N.}\
  \bibnamefont {Hardy}}, \bibinfo {author} {\bibfnamefont {D.~A.}\ \bibnamefont
  {Bonn}}, \bibinfo {author} {\bibfnamefont {Y.}~\bibnamefont {Yoshida}},
  \bibinfo {author} {\bibfnamefont {H.}~\bibnamefont {Eisaki}}, \bibinfo
  {author} {\bibfnamefont {A.~J.}\ \bibnamefont {Achkar}}, \bibinfo {author}
  {\bibfnamefont {D.~G.}\ \bibnamefont {Hawthorn}}, \bibinfo {author}
  {\bibfnamefont {B.}~\bibnamefont {Keimer}}, \bibinfo {author} {\bibfnamefont
  {G.~A.}\ \bibnamefont {Sawatzky}}, \ and\ \bibinfo {author} {\bibfnamefont
  {A.}~\bibnamefont {Damascelli}},\ }\href {\doibase 10.1038/nmat4295}
  {\bibfield  {journal} {\bibinfo  {journal} {Nature Materials}\ }\textbf
  {\bibinfo {volume} {14}},\ \bibinfo {pages} {796} (\bibinfo {year}
  {2015})}\BibitemShut {NoStop}%
\bibitem [{\citenamefont {Hoffman}\ \emph {et~al.}(2002)\citenamefont
  {Hoffman}, \citenamefont {Hudson}, \citenamefont {Lang}, \citenamefont
  {Madhavan}, \citenamefont {Eisaki}, \citenamefont {Uchida},\ and\
  \citenamefont {Davis}}]{Hoffman02}%
  \BibitemOpen
  \bibfield  {author} {\bibinfo {author} {\bibfnamefont {J.~E.}\ \bibnamefont
  {Hoffman}}, \bibinfo {author} {\bibfnamefont {E.~W.}\ \bibnamefont {Hudson}},
  \bibinfo {author} {\bibfnamefont {K.~M.}\ \bibnamefont {Lang}}, \bibinfo
  {author} {\bibfnamefont {V.}~\bibnamefont {Madhavan}}, \bibinfo {author}
  {\bibfnamefont {H.}~\bibnamefont {Eisaki}}, \bibinfo {author} {\bibfnamefont
  {S.}~\bibnamefont {Uchida}}, \ and\ \bibinfo {author} {\bibfnamefont {J.~C.}\
  \bibnamefont {Davis}},\ }\href {\doibase 10.1126/science.1066974} {\bibfield
  {journal} {\bibinfo  {journal} {Science}\ }\textbf {\bibinfo {volume}
  {295}},\ \bibinfo {pages} {466} (\bibinfo {year} {2002})}\BibitemShut
  {NoStop}%
\bibitem [{\citenamefont {da~Silva~Neto}\ \emph {et~al.}(2014)\citenamefont
  {da~Silva~Neto}, \citenamefont {Aynajian}, \citenamefont {Frano},
  \citenamefont {Comin}, \citenamefont {Schierle}, \citenamefont {Weschke},
  \citenamefont {Gyenis}, \citenamefont {Wen}, \citenamefont {Schneeloch},
  \citenamefont {Xu}, \citenamefont {Ono}, \citenamefont {Gu}, \citenamefont
  {Le~Tacon},\ and\ \citenamefont {Yazdani}}]{daSilvaNeto:2014vy}%
  \BibitemOpen
  \bibfield  {author} {\bibinfo {author} {\bibfnamefont {E.~H.}\ \bibnamefont
  {da~Silva~Neto}}, \bibinfo {author} {\bibfnamefont {P.}~\bibnamefont
  {Aynajian}}, \bibinfo {author} {\bibfnamefont {A.}~\bibnamefont {Frano}},
  \bibinfo {author} {\bibfnamefont {R.}~\bibnamefont {Comin}}, \bibinfo
  {author} {\bibfnamefont {E.}~\bibnamefont {Schierle}}, \bibinfo {author}
  {\bibfnamefont {E.}~\bibnamefont {Weschke}}, \bibinfo {author} {\bibfnamefont
  {A.}~\bibnamefont {Gyenis}}, \bibinfo {author} {\bibfnamefont
  {J.}~\bibnamefont {Wen}}, \bibinfo {author} {\bibfnamefont {J.}~\bibnamefont
  {Schneeloch}}, \bibinfo {author} {\bibfnamefont {Z.}~\bibnamefont {Xu}},
  \bibinfo {author} {\bibfnamefont {S.}~\bibnamefont {Ono}}, \bibinfo {author}
  {\bibfnamefont {G.}~\bibnamefont {Gu}}, \bibinfo {author} {\bibfnamefont
  {M.}~\bibnamefont {Le~Tacon}}, \ and\ \bibinfo {author} {\bibfnamefont
  {A.}~\bibnamefont {Yazdani}},\ }\href
  {http://www.sciencemag.org/content/343/6169/393.abstract} {\bibfield
  {journal} {\bibinfo  {journal} {Science}\ }\textbf {\bibinfo {volume}
  {343}},\ \bibinfo {pages} {393} (\bibinfo {year} {2014})}\BibitemShut
  {NoStop}%
\bibitem [{\citenamefont {Mesaros}\ \emph {et~al.}(2016)\citenamefont
  {Mesaros}, \citenamefont {Fujita}, \citenamefont {Edkins}, \citenamefont
  {Hamidian}, \citenamefont {Eisaki}, \citenamefont {Uchida}, \citenamefont
  {Davis}, \citenamefont {Lawler},\ and\ \citenamefont {Kim}}]{Mesaros:2016fe}%
  \BibitemOpen
  \bibfield  {author} {\bibinfo {author} {\bibfnamefont {A.}~\bibnamefont
  {Mesaros}}, \bibinfo {author} {\bibfnamefont {K.}~\bibnamefont {Fujita}},
  \bibinfo {author} {\bibfnamefont {S.~D.}\ \bibnamefont {Edkins}}, \bibinfo
  {author} {\bibfnamefont {M.~H.}\ \bibnamefont {Hamidian}}, \bibinfo {author}
  {\bibfnamefont {H.}~\bibnamefont {Eisaki}}, \bibinfo {author} {\bibfnamefont
  {S.-i.}\ \bibnamefont {Uchida}}, \bibinfo {author} {\bibfnamefont {J.~C.~S.}\
  \bibnamefont {Davis}}, \bibinfo {author} {\bibfnamefont {M.~J.}\ \bibnamefont
  {Lawler}}, \ and\ \bibinfo {author} {\bibfnamefont {E.-A.}\ \bibnamefont
  {Kim}},\ }\href {\doibase 10.1073/pnas.1614247113} {\bibfield  {journal}
  {\bibinfo  {journal} {Proc. Natl. Acad. Sci.}\ }\textbf {\bibinfo {volume}
  {113}},\ \bibinfo {pages} {12661} (\bibinfo {year} {2016})}\BibitemShut
  {NoStop}%
\bibitem [{\citenamefont {Wu}\ \emph {et~al.}(2015)\citenamefont {Wu},
  \citenamefont {Mayaffre}, \citenamefont {Kr{\"a}mer},\ and\ \citenamefont
  {Horvati{\'c}}}]{Wu:2015bt}%
  \BibitemOpen
  \bibfield  {author} {\bibinfo {author} {\bibfnamefont {T.}~\bibnamefont
  {Wu}}, \bibinfo {author} {\bibfnamefont {H.}~\bibnamefont {Mayaffre}},
  \bibinfo {author} {\bibfnamefont {S.}~\bibnamefont {Kr{\"a}mer}}, \ and\
  \bibinfo {author} {\bibfnamefont {M.}~\bibnamefont {Horvati{\'c}}},\ }\href
  {\doibase 10.1038/ncomms7438} {\bibfield  {journal} {\bibinfo  {journal}
  {Nature}\ }\textbf {\bibinfo {volume} {6}},\ \bibinfo {pages} {6438}
  (\bibinfo {year} {2015})}\BibitemShut {NoStop}%
\bibitem [{\citenamefont {{Hamidian}}\ \emph {et~al.}(2015)\citenamefont
  {{Hamidian}}, \citenamefont {{Edkins}}, \citenamefont {{Fujita}},
  \citenamefont {{Mackenzie}}, \citenamefont {{Eisaki}}, \citenamefont
  {{Uchida}}, \citenamefont {{Lawler}}, \citenamefont {{Kim}}, \citenamefont
  {{Sachdev}},\ and\ \citenamefont {{S{\'e}amus Davis}}}]{Hamidian15}%
  \BibitemOpen
  \bibfield  {author} {\bibinfo {author} {\bibfnamefont {M.~H.}\ \bibnamefont
  {{Hamidian}}}, \bibinfo {author} {\bibfnamefont {S.~D.}\ \bibnamefont
  {{Edkins}}}, \bibinfo {author} {\bibfnamefont {K.}~\bibnamefont {{Fujita}}},
  \bibinfo {author} {\bibfnamefont {A.~P.}\ \bibnamefont {{Mackenzie}}},
  \bibinfo {author} {\bibfnamefont {H.}~\bibnamefont {{Eisaki}}}, \bibinfo
  {author} {\bibfnamefont {S.}~\bibnamefont {{Uchida}}}, \bibinfo {author}
  {\bibfnamefont {M.~J.}\ \bibnamefont {{Lawler}}}, \bibinfo {author}
  {\bibfnamefont {E.-A.}\ \bibnamefont {{Kim}}}, \bibinfo {author}
  {\bibfnamefont {S.}~\bibnamefont {{Sachdev}}}, \ and\ \bibinfo {author}
  {\bibfnamefont {J.~C.}\ \bibnamefont {{S{\'e}amus Davis}}},\ }\href
  {http://arxiv.org/abs/1508.00620} {\  (\bibinfo {year} {2015})},\ \Eprint
  {http://arxiv.org/abs/1508.00620} {arXiv:1508.00620} \BibitemShut {NoStop}%
\bibitem [{\citenamefont {Hamidian}\ \emph {et~al.}(2015)\citenamefont
  {Hamidian}, \citenamefont {Edkins}, \citenamefont {Kim}, \citenamefont
  {Davis}, \citenamefont {Mackenzie}, \citenamefont {Eisaki}, \citenamefont
  {Uchida}, \citenamefont {Lawler}, \citenamefont {Kim}, \citenamefont
  {Sachdev},\ and\ \citenamefont {Fujita}}]{Hamidian15a}%
  \BibitemOpen
  \bibfield  {author} {\bibinfo {author} {\bibfnamefont {M.~H.}\ \bibnamefont
  {Hamidian}}, \bibinfo {author} {\bibfnamefont {S.~D.}\ \bibnamefont
  {Edkins}}, \bibinfo {author} {\bibfnamefont {C.~K.}\ \bibnamefont {Kim}},
  \bibinfo {author} {\bibfnamefont {J.~C.}\ \bibnamefont {Davis}}, \bibinfo
  {author} {\bibfnamefont {A.~P.}\ \bibnamefont {Mackenzie}}, \bibinfo {author}
  {\bibfnamefont {H.}~\bibnamefont {Eisaki}}, \bibinfo {author} {\bibfnamefont
  {S.}~\bibnamefont {Uchida}}, \bibinfo {author} {\bibfnamefont {M.~J.}\
  \bibnamefont {Lawler}}, \bibinfo {author} {\bibfnamefont {E.-A.}\
  \bibnamefont {Kim}}, \bibinfo {author} {\bibfnamefont {S.}~\bibnamefont
  {Sachdev}}, \ and\ \bibinfo {author} {\bibfnamefont {K.}~\bibnamefont
  {Fujita}},\ }\href {http://dx.doi.org/10.1038/nphys3519} {\bibfield
  {journal} {\bibinfo  {journal} {Nat. Phys.}\ }\textbf {\bibinfo {volume}
  {12}},\ \bibinfo {pages} {150} (\bibinfo {year} {2015})}\BibitemShut
  {NoStop}%
\bibitem [{\citenamefont {Montiel}\ \emph {et~al.}(2017)\citenamefont
  {Montiel}, \citenamefont {Kloss},\ and\ \citenamefont {Pepin}}]{Montiel16}%
  \BibitemOpen
  \bibfield  {author} {\bibinfo {author} {\bibfnamefont {X.}~\bibnamefont
  {Montiel}}, \bibinfo {author} {\bibfnamefont {T.}~\bibnamefont {Kloss}}, \
  and\ \bibinfo {author} {\bibfnamefont {C.}~\bibnamefont {Pepin}},\
  }\href@noop {} {\bibfield  {journal} {\bibinfo  {journal} {Phys. Rev. B}\
  }\textbf {\bibinfo {volume} {95}},\ \bibinfo {pages} {104510} (\bibinfo
  {year} {2017})}\BibitemShut {NoStop}%
\bibitem [{\citenamefont {Fauqu\'e}\ \emph {et~al.}(2006)\citenamefont
  {Fauqu\'e}, \citenamefont {Sidis}, \citenamefont {Hinkov}, \citenamefont
  {Pailh\`es}, \citenamefont {Lin}, \citenamefont {Chaud},\ and\ \citenamefont
  {Bourges}}]{Fauque06}%
  \BibitemOpen
  \bibfield  {author} {\bibinfo {author} {\bibfnamefont {B.}~\bibnamefont
  {Fauqu\'e}}, \bibinfo {author} {\bibfnamefont {Y.}~\bibnamefont {Sidis}},
  \bibinfo {author} {\bibfnamefont {V.}~\bibnamefont {Hinkov}}, \bibinfo
  {author} {\bibfnamefont {S.}~\bibnamefont {Pailh\`es}}, \bibinfo {author}
  {\bibfnamefont {C.~T.}\ \bibnamefont {Lin}}, \bibinfo {author} {\bibfnamefont
  {X.}~\bibnamefont {Chaud}}, \ and\ \bibinfo {author} {\bibfnamefont
  {P.}~\bibnamefont {Bourges}},\ }\href {\doibase
  10.1103/PhysRevLett.96.197001} {\bibfield  {journal} {\bibinfo  {journal}
  {Phys. Rev. Lett.}\ }\textbf {\bibinfo {volume} {96}},\ \bibinfo {pages}
  {197001} (\bibinfo {year} {2006})}\BibitemShut {NoStop}%
\bibitem [{\citenamefont {Mangin-Thro}\ \emph {et~al.}(2014)\citenamefont
  {Mangin-Thro}, \citenamefont {Sidis},\ and\ \citenamefont
  {Bourges}}]{ManginThro:2014js}%
  \BibitemOpen
  \bibfield  {author} {\bibinfo {author} {\bibfnamefont {L.}~\bibnamefont
  {Mangin-Thro}}, \bibinfo {author} {\bibfnamefont {Y.}~\bibnamefont {Sidis}},
  \ and\ \bibinfo {author} {\bibfnamefont {P.}~\bibnamefont {Bourges}},\ }\href
  {\doibase 10.1103/physrevb.89.094523} {\bibfield  {journal} {\bibinfo
  {journal} {Phys. Rev. B}\ }\textbf {\bibinfo {volume} {89}},\ \bibinfo
  {pages} {094523} (\bibinfo {year} {2014})}\BibitemShut {NoStop}%
\bibitem [{\citenamefont {Mangin-Thro}\ \emph {et~al.}(2015)\citenamefont
  {Mangin-Thro}, \citenamefont {Sidis}, \citenamefont {Wildes},\ and\
  \citenamefont {Bourges}}]{ManginThro:2015fg}%
  \BibitemOpen
  \bibfield  {author} {\bibinfo {author} {\bibfnamefont {L.}~\bibnamefont
  {Mangin-Thro}}, \bibinfo {author} {\bibfnamefont {Y.}~\bibnamefont {Sidis}},
  \bibinfo {author} {\bibfnamefont {A.}~\bibnamefont {Wildes}}, \ and\ \bibinfo
  {author} {\bibfnamefont {P.}~\bibnamefont {Bourges}},\ }\href {\doibase
  10.1038/ncomms8705} {\bibfield  {journal} {\bibinfo  {journal} {Nat.
  Commun.}\ }\textbf {\bibinfo {volume} {6}},\ \bibinfo {pages} {7705}
  (\bibinfo {year} {2015})}\BibitemShut {NoStop}%
\bibitem [{\citenamefont {Mesaros}\ \emph {et~al.}(2011)\citenamefont
  {Mesaros}, \citenamefont {Fujita}, \citenamefont {Eisaki}, \citenamefont
  {Uchida},\ and\ \citenamefont {Davis}}]{Mesaros:2011jj}%
  \BibitemOpen
  \bibfield  {author} {\bibinfo {author} {\bibfnamefont {A.}~\bibnamefont
  {Mesaros}}, \bibinfo {author} {\bibfnamefont {K.}~\bibnamefont {Fujita}},
  \bibinfo {author} {\bibfnamefont {H.}~\bibnamefont {Eisaki}}, \bibinfo
  {author} {\bibfnamefont {S.}~\bibnamefont {Uchida}}, \ and\ \bibinfo {author}
  {\bibfnamefont {J.~C.}\ \bibnamefont {Davis}},\ }\href {\doibase
  10.1126/science.1201082} {\bibfield  {journal} {\bibinfo  {journal}
  {Science}\ }\textbf {\bibinfo {volume} {333}},\ \bibinfo {pages} {426}
  (\bibinfo {year} {2011})}\BibitemShut {NoStop}%
\bibitem [{\citenamefont {{Sato}}\ \emph {et~al.}(2017)\citenamefont {{Sato}},
  \citenamefont {{Kasahara}}, \citenamefont {{Murayama}}, \citenamefont
  {{Kasahara}}, \citenamefont {{Moon}}, \citenamefont {{Nishizaki}},
  \citenamefont {{Loew}}, \citenamefont {{Porras}}, \citenamefont {{Keimer}},
  \citenamefont {{Shibauchi}},\ and\ \citenamefont {{Matsuda}}}]{Sato2017}%
  \BibitemOpen
  \bibfield  {author} {\bibinfo {author} {\bibfnamefont {Y.}~\bibnamefont
  {{Sato}}}, \bibinfo {author} {\bibfnamefont {S.}~\bibnamefont {{Kasahara}}},
  \bibinfo {author} {\bibfnamefont {H.}~\bibnamefont {{Murayama}}}, \bibinfo
  {author} {\bibfnamefont {Y.}~\bibnamefont {{Kasahara}}}, \bibinfo {author}
  {\bibfnamefont {E.-G.}\ \bibnamefont {{Moon}}}, \bibinfo {author}
  {\bibfnamefont {T.}~\bibnamefont {{Nishizaki}}}, \bibinfo {author}
  {\bibfnamefont {T.}~\bibnamefont {{Loew}}}, \bibinfo {author} {\bibfnamefont
  {J.}~\bibnamefont {{Porras}}}, \bibinfo {author} {\bibfnamefont
  {B.}~\bibnamefont {{Keimer}}}, \bibinfo {author} {\bibfnamefont
  {T.}~\bibnamefont {{Shibauchi}}}, \ and\ \bibinfo {author} {\bibfnamefont
  {Y.}~\bibnamefont {{Matsuda}}},\ }\href {\doibase 10.1038/nphys4205}
  {\bibfield  {journal} {\bibinfo  {journal} {Nat. Phys.}\ }\textbf {\bibinfo
  {volume} {13}},\ \bibinfo {pages} {1074} (\bibinfo {year}
  {2017})}\BibitemShut {NoStop}%
\bibitem [{\citenamefont {Gomes}\ \emph {et~al.}(2007)\citenamefont {Gomes},
  \citenamefont {Pasupathy}, \citenamefont {Pushp}, \citenamefont {Ono},\ and\
  \citenamefont {Ando}}]{Gomes:2007ks}%
  \BibitemOpen
  \bibfield  {author} {\bibinfo {author} {\bibfnamefont {K.~K.}\ \bibnamefont
  {Gomes}}, \bibinfo {author} {\bibfnamefont {A.~N.}\ \bibnamefont
  {Pasupathy}}, \bibinfo {author} {\bibfnamefont {A.}~\bibnamefont {Pushp}},
  \bibinfo {author} {\bibfnamefont {S.}~\bibnamefont {Ono}}, \ and\ \bibinfo
  {author} {\bibfnamefont {Y.}~\bibnamefont {Ando}},\ }\href {\doibase
  10.1038/nature05881} {\bibfield  {journal} {\bibinfo  {journal} {Nature}\
  }\textbf {\bibinfo {volume} {447}},\ \bibinfo {pages} {569} (\bibinfo {year}
  {2007})}\BibitemShut {NoStop}%
\bibitem [{\citenamefont {Kohsaka}\ \emph {et~al.}(2007)\citenamefont
  {Kohsaka}, \citenamefont {Taylor}, \citenamefont {Fujita}, \citenamefont
  {Schmidt}, \citenamefont {Lupien}, \citenamefont {Hanaguri}, \citenamefont
  {Azuma}, \citenamefont {Takano}, \citenamefont {Eisaki}, \citenamefont
  {Takagi}, \citenamefont {Uchida},\ and\ \citenamefont {Davis}}]{Kohsaka07}%
  \BibitemOpen
  \bibfield  {author} {\bibinfo {author} {\bibfnamefont {Y.}~\bibnamefont
  {Kohsaka}}, \bibinfo {author} {\bibfnamefont {C.}~\bibnamefont {Taylor}},
  \bibinfo {author} {\bibfnamefont {K.}~\bibnamefont {Fujita}}, \bibinfo
  {author} {\bibfnamefont {A.}~\bibnamefont {Schmidt}}, \bibinfo {author}
  {\bibfnamefont {C.}~\bibnamefont {Lupien}}, \bibinfo {author} {\bibfnamefont
  {T.}~\bibnamefont {Hanaguri}}, \bibinfo {author} {\bibfnamefont
  {M.}~\bibnamefont {Azuma}}, \bibinfo {author} {\bibfnamefont
  {M.}~\bibnamefont {Takano}}, \bibinfo {author} {\bibfnamefont
  {H.}~\bibnamefont {Eisaki}}, \bibinfo {author} {\bibfnamefont
  {H.}~\bibnamefont {Takagi}}, \bibinfo {author} {\bibfnamefont
  {S.}~\bibnamefont {Uchida}}, \ and\ \bibinfo {author} {\bibfnamefont {J.~C.}\
  \bibnamefont {Davis}},\ }\href {\doibase 10.1126/science.1138584} {\bibfield
  {journal} {\bibinfo  {journal} {Science}\ }\textbf {\bibinfo {volume}
  {315}},\ \bibinfo {pages} {1380} (\bibinfo {year} {2007})}\BibitemShut
  {NoStop}%
\bibitem [{\citenamefont {Fradkin}\ \emph {et~al.}(2015)\citenamefont
  {Fradkin}, \citenamefont {Kivelson},\ and\ \citenamefont
  {Tranquada}}]{Fradkin:2015ch}%
  \BibitemOpen
  \bibfield  {author} {\bibinfo {author} {\bibfnamefont {E.}~\bibnamefont
  {Fradkin}}, \bibinfo {author} {\bibfnamefont {S.~A.}\ \bibnamefont
  {Kivelson}}, \ and\ \bibinfo {author} {\bibfnamefont {J.~M.}\ \bibnamefont
  {Tranquada}},\ }\href {\doibase 10.1103/revmodphys.87.457} {\bibfield
  {journal} {\bibinfo  {journal} {Rev. Mod. Phys.}\ }\textbf {\bibinfo {volume}
  {87}},\ \bibinfo {pages} {457} (\bibinfo {year} {2015})}\BibitemShut
  {NoStop}%
\bibitem [{\citenamefont {Metlitski}\ and\ \citenamefont
  {Sachdev}(2010)}]{Metlitski10}%
  \BibitemOpen
  \bibfield  {author} {\bibinfo {author} {\bibfnamefont {M.~A.}\ \bibnamefont
  {Metlitski}}\ and\ \bibinfo {author} {\bibfnamefont {S.}~\bibnamefont
  {Sachdev}},\ }\href {http://stacks.iop.org/1367-2630/12/i=10/a=105007}
  {\bibfield  {journal} {\bibinfo  {journal} {New J. Phys.}\ }\textbf {\bibinfo
  {volume} {12}},\ \bibinfo {pages} {105007} (\bibinfo {year}
  {2010})}\BibitemShut {NoStop}%
\bibitem [{\citenamefont {Efetov}\ \emph {et~al.}(2013)\citenamefont {Efetov},
  \citenamefont {Meier},\ and\ \citenamefont {P\'epin}}]{Efetov13}%
  \BibitemOpen
  \bibfield  {author} {\bibinfo {author} {\bibfnamefont {K.~B.}\ \bibnamefont
  {Efetov}}, \bibinfo {author} {\bibfnamefont {H.}~\bibnamefont {Meier}}, \
  and\ \bibinfo {author} {\bibfnamefont {C.}~\bibnamefont {P\'epin}},\ }\href
  {\doibase 10.1038/nphys2641} {\bibfield  {journal} {\bibinfo  {journal} {Nat.
  Phys.}\ }\textbf {\bibinfo {volume} {9}},\ \bibinfo {pages} {442} (\bibinfo
  {year} {2013})}\BibitemShut {NoStop}%
\bibitem [{\citenamefont {Kloss}\ \emph {et~al.}(2015)\citenamefont {Kloss},
  \citenamefont {Montiel},\ and\ \citenamefont {P\'epin}}]{Kloss15}%
  \BibitemOpen
  \bibfield  {author} {\bibinfo {author} {\bibfnamefont {T.}~\bibnamefont
  {Kloss}}, \bibinfo {author} {\bibfnamefont {X.}~\bibnamefont {Montiel}}, \
  and\ \bibinfo {author} {\bibfnamefont {C.}~\bibnamefont {P\'epin}},\ }\href
  {\doibase 10.1103/PhysRevB.91.205124} {\bibfield  {journal} {\bibinfo
  {journal} {Phys. Rev. B}\ }\textbf {\bibinfo {volume} {91}},\ \bibinfo
  {pages} {205124} (\bibinfo {year} {2015})}\BibitemShut {NoStop}%
\bibitem [{\citenamefont {Mishra}\ and\ \citenamefont
  {Norman}(2015)}]{Mishra2015}%
  \BibitemOpen
  \bibfield  {author} {\bibinfo {author} {\bibfnamefont {V.}~\bibnamefont
  {Mishra}}\ and\ \bibinfo {author} {\bibfnamefont {M.~R.}\ \bibnamefont
  {Norman}},\ }\href {\doibase 10.1103/PhysRevB.92.060507} {\bibfield
  {journal} {\bibinfo  {journal} {Phys. Rev. B}\ }\textbf {\bibinfo {volume}
  {92}},\ \bibinfo {pages} {060507} (\bibinfo {year} {2015})}\BibitemShut
  {NoStop}%
\bibitem [{\citenamefont {Wang}\ and\ \citenamefont {Chubukov}(2014)}]{Wang14}%
  \BibitemOpen
  \bibfield  {author} {\bibinfo {author} {\bibfnamefont {Y.}~\bibnamefont
  {Wang}}\ and\ \bibinfo {author} {\bibfnamefont {A.}~\bibnamefont
  {Chubukov}},\ }\href {\doibase 10.1103/PhysRevB.90.035149} {\bibfield
  {journal} {\bibinfo  {journal} {Phys. Rev. B}\ }\textbf {\bibinfo {volume}
  {90}},\ \bibinfo {pages} {035149} (\bibinfo {year} {2014})}\BibitemShut
  {NoStop}%
\bibitem [{\citenamefont {{Wang}}\ and\ \citenamefont
  {{Chubukov}}(2015)}]{WangReply}%
  \BibitemOpen
  \bibfield  {author} {\bibinfo {author} {\bibfnamefont {Y.}~\bibnamefont
  {{Wang}}}\ and\ \bibinfo {author} {\bibfnamefont {A.}~\bibnamefont
  {{Chubukov}}},\ }\href@noop {} {\bibfield  {journal} {\bibinfo  {journal}
  {ArXiv e-prints}\ } (\bibinfo {year} {2015})},\ \Eprint
  {http://arxiv.org/abs/1502.07689} {arXiv:1502.07689} \BibitemShut {NoStop}%
\bibitem [{\citenamefont {Villain}\ \emph {et~al.}(1980)\citenamefont
  {Villain}, \citenamefont {Bidaux}, \citenamefont {Carton},\ and\
  \citenamefont {R.}}]{Villain80}%
  \BibitemOpen
  \bibfield  {author} {\bibinfo {author} {\bibfnamefont {J.}~\bibnamefont
  {Villain}}, \bibinfo {author} {\bibfnamefont {R.}~\bibnamefont {Bidaux}},
  \bibinfo {author} {\bibfnamefont {J.~P.}\ \bibnamefont {Carton}}, \ and\
  \bibinfo {author} {\bibfnamefont {C.}~\bibnamefont {R.}},\ }\href@noop {}
  {\bibfield  {journal} {\bibinfo  {journal} {J. Physique}\ }\textbf {\bibinfo
  {volume} {41}},\ \bibinfo {pages} {1263} (\bibinfo {year}
  {1980})}\BibitemShut {NoStop}%
\bibitem [{\citenamefont {Mermin}(1979)}]{Mermin:1979io}%
  \BibitemOpen
  \bibfield  {author} {\bibinfo {author} {\bibfnamefont {N.~D.}\ \bibnamefont
  {Mermin}},\ }\href {\doibase 10.1103/RevModPhys.51.591} {\bibfield  {journal}
  {\bibinfo  {journal} {Rev. Mod. Phys.}\ }\textbf {\bibinfo {volume} {51}},\
  \bibinfo {pages} {591} (\bibinfo {year} {1979})}\BibitemShut {NoStop}%
\bibitem [{\citenamefont {Lee}\ and\ \citenamefont {Rice}(1979)}]{Lee:1979hu}%
  \BibitemOpen
  \bibfield  {author} {\bibinfo {author} {\bibfnamefont {P.~A.}\ \bibnamefont
  {Lee}}\ and\ \bibinfo {author} {\bibfnamefont {T.~M.}\ \bibnamefont {Rice}},\
  }\href {\doibase 10.1103/PhysRevB.19.3970} {\bibfield  {journal} {\bibinfo
  {journal} {Phys. Rev. B}\ }\textbf {\bibinfo {volume} {19}},\ \bibinfo
  {pages} {3970} (\bibinfo {year} {1979})}\BibitemShut {NoStop}%
\bibitem [{\citenamefont {Lee}\ and\ \citenamefont {Wen}(2001)}]{Lee:2001ib}%
  \BibitemOpen
  \bibfield  {author} {\bibinfo {author} {\bibfnamefont {P.~A.}\ \bibnamefont
  {Lee}}\ and\ \bibinfo {author} {\bibfnamefont {X.-G.}\ \bibnamefont {Wen}},\
  }\href {\doibase 10.1103/PhysRevB.63.224517} {\bibfield  {journal} {\bibinfo
  {journal} {Phys. Rev. B}\ }\textbf {\bibinfo {volume} {63}},\ \bibinfo
  {pages} {224517} (\bibinfo {year} {2001})}\BibitemShut {NoStop}%
\bibitem [{\citenamefont {Lee}\ \emph {et~al.}(2006)\citenamefont {Lee},
  \citenamefont {Nagaosa},\ and\ \citenamefont {Wen}}]{Lee06}%
  \BibitemOpen
  \bibfield  {author} {\bibinfo {author} {\bibfnamefont {P.~A.}\ \bibnamefont
  {Lee}}, \bibinfo {author} {\bibfnamefont {N.}~\bibnamefont {Nagaosa}}, \ and\
  \bibinfo {author} {\bibfnamefont {X.-G.}\ \bibnamefont {Wen}},\ }\href
  {\doibase 10.1103/RevModPhys.78.17} {\bibfield  {journal} {\bibinfo
  {journal} {Rev. Mod. Phys.}\ }\textbf {\bibinfo {volume} {78}},\ \bibinfo
  {pages} {17} (\bibinfo {year} {2006})}\BibitemShut {NoStop}%
\bibitem [{\citenamefont {Sheehy}\ and\ \citenamefont
  {Goldbart}(1998)}]{Sheehy1998}%
  \BibitemOpen
  \bibfield  {author} {\bibinfo {author} {\bibfnamefont {D.~E.}\ \bibnamefont
  {Sheehy}}\ and\ \bibinfo {author} {\bibfnamefont {P.~M.}\ \bibnamefont
  {Goldbart}},\ }\href {\doibase 10.1103/PhysRevB.57.R8131} {\bibfield
  {journal} {\bibinfo  {journal} {Phys. Rev. B}\ }\textbf {\bibinfo {volume}
  {57}},\ \bibinfo {pages} {R8131} (\bibinfo {year} {1998})}\BibitemShut
  {NoStop}%
\bibitem [{\citenamefont {Goldbart}\ and\ \citenamefont
  {Sheehy}(1998)}]{Goldbart1998}%
  \BibitemOpen
  \bibfield  {author} {\bibinfo {author} {\bibfnamefont {P.~M.}\ \bibnamefont
  {Goldbart}}\ and\ \bibinfo {author} {\bibfnamefont {D.~E.}\ \bibnamefont
  {Sheehy}},\ }\href {\doibase 10.1103/PhysRevB.58.5731} {\bibfield  {journal}
  {\bibinfo  {journal} {Phys. Rev. B}\ }\textbf {\bibinfo {volume} {58}},\
  \bibinfo {pages} {5731} (\bibinfo {year} {1998})}\BibitemShut {NoStop}%
\bibitem [{\citenamefont {Muehlbauer}\ \emph {et~al.}(2009)\citenamefont
  {Muehlbauer}, \citenamefont {Binz}, \citenamefont {Jonietz}, \citenamefont
  {Pfleiderer}, \citenamefont {Rosch}, \citenamefont {Neubauer}, \citenamefont
  {Georgii},\ and\ \citenamefont {Boeni}}]{Muehlbauer:2009bc}%
  \BibitemOpen
  \bibfield  {author} {\bibinfo {author} {\bibfnamefont {S.}~\bibnamefont
  {Muehlbauer}}, \bibinfo {author} {\bibfnamefont {B.}~\bibnamefont {Binz}},
  \bibinfo {author} {\bibfnamefont {F.}~\bibnamefont {Jonietz}}, \bibinfo
  {author} {\bibfnamefont {C.}~\bibnamefont {Pfleiderer}}, \bibinfo {author}
  {\bibfnamefont {A.}~\bibnamefont {Rosch}}, \bibinfo {author} {\bibfnamefont
  {A.}~\bibnamefont {Neubauer}}, \bibinfo {author} {\bibfnamefont
  {R.}~\bibnamefont {Georgii}}, \ and\ \bibinfo {author} {\bibfnamefont
  {P.}~\bibnamefont {Boeni}},\ }\href {\doibase 10.1126/science.1166767}
  {\bibfield  {journal} {\bibinfo  {journal} {Science}\ }\textbf {\bibinfo
  {volume} {323}},\ \bibinfo {pages} {915} (\bibinfo {year}
  {2009})}\BibitemShut {NoStop}%
\bibitem [{\citenamefont {Romming}\ \emph {et~al.}(2013)\citenamefont
  {Romming}, \citenamefont {Hanneken}, \citenamefont {Menzel}, \citenamefont
  {Bickel}, \citenamefont {Wolter}, \citenamefont {von Bergmann}, \citenamefont
  {Kubetzka},\ and\ \citenamefont {Wiesendanger}}]{Romming636}%
  \BibitemOpen
  \bibfield  {author} {\bibinfo {author} {\bibfnamefont {N.}~\bibnamefont
  {Romming}}, \bibinfo {author} {\bibfnamefont {C.}~\bibnamefont {Hanneken}},
  \bibinfo {author} {\bibfnamefont {M.}~\bibnamefont {Menzel}}, \bibinfo
  {author} {\bibfnamefont {J.~E.}\ \bibnamefont {Bickel}}, \bibinfo {author}
  {\bibfnamefont {B.}~\bibnamefont {Wolter}}, \bibinfo {author} {\bibfnamefont
  {K.}~\bibnamefont {von Bergmann}}, \bibinfo {author} {\bibfnamefont
  {A.}~\bibnamefont {Kubetzka}}, \ and\ \bibinfo {author} {\bibfnamefont
  {R.}~\bibnamefont {Wiesendanger}},\ }\href {\doibase 10.1126/science.1240573}
  {\bibfield  {journal} {\bibinfo  {journal} {Science}\ }\textbf {\bibinfo
  {volume} {341}},\ \bibinfo {pages} {636} (\bibinfo {year}
  {2013})}\BibitemShut {NoStop}%
\bibitem [{\citenamefont {Wu}\ \emph {et~al.}(2013)\citenamefont {Wu},
  \citenamefont {Mayaffre}, \citenamefont {Kr{\"a}mer}, \citenamefont
  {Horvati{\'c}}, \citenamefont {Berthier}, \citenamefont {Kuhns},
  \citenamefont {Reyes}, \citenamefont {Liang}, \citenamefont {Hardy},
  \citenamefont {Bonn},\ and\ \citenamefont {Julien}}]{Wu13a}%
  \BibitemOpen
  \bibfield  {author} {\bibinfo {author} {\bibfnamefont {T.}~\bibnamefont
  {Wu}}, \bibinfo {author} {\bibfnamefont {H.}~\bibnamefont {Mayaffre}},
  \bibinfo {author} {\bibfnamefont {S.}~\bibnamefont {Kr{\"a}mer}}, \bibinfo
  {author} {\bibfnamefont {M.}~\bibnamefont {Horvati{\'c}}}, \bibinfo {author}
  {\bibfnamefont {C.}~\bibnamefont {Berthier}}, \bibinfo {author}
  {\bibfnamefont {P.~L.}\ \bibnamefont {Kuhns}}, \bibinfo {author}
  {\bibfnamefont {A.~P.}\ \bibnamefont {Reyes}}, \bibinfo {author}
  {\bibfnamefont {R.}~\bibnamefont {Liang}}, \bibinfo {author} {\bibfnamefont
  {W.~N.}\ \bibnamefont {Hardy}}, \bibinfo {author} {\bibfnamefont {D.~A.}\
  \bibnamefont {Bonn}}, \ and\ \bibinfo {author} {\bibfnamefont {M.-H.}\
  \bibnamefont {Julien}},\ }\href {\doibase 10.1038/ncomms3113} {\bibfield
  {journal} {\bibinfo  {journal} {Nat. Commun.}\ }\textbf {\bibinfo {volume}
  {4}},\ \bibinfo {pages} {2113} (\bibinfo {year} {2013})}\BibitemShut
  {NoStop}%
\bibitem [{\citenamefont {Wang}\ \emph {et~al.}(2002)\citenamefont {Wang},
  \citenamefont {Ong}, \citenamefont {Xu}, \citenamefont {Kakeshita},
  \citenamefont {Uchida}, \citenamefont {Bonn}, \citenamefont {Liang},\ and\
  \citenamefont {Hardy}}]{Wang:2002ke}%
  \BibitemOpen
  \bibfield  {author} {\bibinfo {author} {\bibfnamefont {Y.}~\bibnamefont
  {Wang}}, \bibinfo {author} {\bibfnamefont {N.~P.}\ \bibnamefont {Ong}},
  \bibinfo {author} {\bibfnamefont {Z.~A.}\ \bibnamefont {Xu}}, \bibinfo
  {author} {\bibfnamefont {T.}~\bibnamefont {Kakeshita}}, \bibinfo {author}
  {\bibfnamefont {S.}~\bibnamefont {Uchida}}, \bibinfo {author} {\bibfnamefont
  {D.~A.}\ \bibnamefont {Bonn}}, \bibinfo {author} {\bibfnamefont
  {R.}~\bibnamefont {Liang}}, \ and\ \bibinfo {author} {\bibfnamefont {W.~N.}\
  \bibnamefont {Hardy}},\ }\href {\doibase 10.1103/physrevlett.88.257003}
  {\bibfield  {journal} {\bibinfo  {journal} {Phys. Rev. Lett.}\ }\textbf
  {\bibinfo {volume} {88}},\ \bibinfo {pages} {257003} (\bibinfo {year}
  {2002})}\BibitemShut {NoStop}%
\bibitem [{\citenamefont {Alloul}\ \emph {et~al.}(2010)\citenamefont {Alloul},
  \citenamefont {Rullier-Albenque}, \citenamefont {Vignolle}, \citenamefont
  {Colson},\ and\ \citenamefont {Forget}}]{Alloul:2010ko}%
  \BibitemOpen
  \bibfield  {author} {\bibinfo {author} {\bibfnamefont {H.}~\bibnamefont
  {Alloul}}, \bibinfo {author} {\bibfnamefont {F.}~\bibnamefont
  {Rullier-Albenque}}, \bibinfo {author} {\bibfnamefont {B.}~\bibnamefont
  {Vignolle}}, \bibinfo {author} {\bibfnamefont {D.}~\bibnamefont {Colson}}, \
  and\ \bibinfo {author} {\bibfnamefont {A.}~\bibnamefont {Forget}},\ }\href
  {\doibase 10.1209/0295-5075/91/37005} {\bibfield  {journal} {\bibinfo
  {journal} {Europhys. Lett.}\ }\textbf {\bibinfo {volume} {91}},\ \bibinfo
  {pages} {37005} (\bibinfo {year} {2010})}\BibitemShut {NoStop}%
\bibitem [{\citenamefont {Garate}(2013)}]{Garate2013}%
  \BibitemOpen
  \bibfield  {author} {\bibinfo {author} {\bibfnamefont {I.}~\bibnamefont
  {Garate}},\ }\href {\doibase 10.1103/PhysRevB.88.094511} {\bibfield
  {journal} {\bibinfo  {journal} {Phys. Rev. B}\ }\textbf {\bibinfo {volume}
  {88}},\ \bibinfo {pages} {094511} (\bibinfo {year} {2013})}\BibitemShut
  {NoStop}%
\bibitem [{\citenamefont {Cyr-Choiniere}\ \emph {et~al.}(2015)\citenamefont
  {Cyr-Choiniere}, \citenamefont {Grissonnanche}, \citenamefont {Badoux},
  \citenamefont {Day}, \citenamefont {Bonn}, \citenamefont {Hardy},
  \citenamefont {Liang}, \citenamefont {Doiron-Leyraud},\ and\ \citenamefont
  {Taillefer}}]{CyrChoiniere:2015dk}%
  \BibitemOpen
  \bibfield  {author} {\bibinfo {author} {\bibfnamefont {O.}~\bibnamefont
  {Cyr-Choiniere}}, \bibinfo {author} {\bibfnamefont {G.}~\bibnamefont
  {Grissonnanche}}, \bibinfo {author} {\bibfnamefont {S.}~\bibnamefont
  {Badoux}}, \bibinfo {author} {\bibfnamefont {J.}~\bibnamefont {Day}},
  \bibinfo {author} {\bibfnamefont {D.~A.}\ \bibnamefont {Bonn}}, \bibinfo
  {author} {\bibfnamefont {W.~N.}\ \bibnamefont {Hardy}}, \bibinfo {author}
  {\bibfnamefont {R.}~\bibnamefont {Liang}}, \bibinfo {author} {\bibfnamefont
  {N.}~\bibnamefont {Doiron-Leyraud}}, \ and\ \bibinfo {author} {\bibfnamefont
  {L.}~\bibnamefont {Taillefer}},\ }\href {\doibase 10.1103/PhysRevB.92.224502}
  {\bibfield  {journal} {\bibinfo  {journal} {Phys. Rev. B}\ }\textbf {\bibinfo
  {volume} {92}},\ \bibinfo {pages} {224502} (\bibinfo {year}
  {2015})}\BibitemShut {NoStop}%
\bibitem [{\citenamefont {Shekhter}\ \emph {et~al.}(2013)\citenamefont
  {Shekhter}, \citenamefont {Ramshaw}, \citenamefont {Liang}, \citenamefont
  {Hardy}, \citenamefont {Bonn}, \citenamefont {Balakirev}, \citenamefont
  {McDonald}, \citenamefont {Betts}, \citenamefont {Riggs},\ and\ \citenamefont
  {Migliori}}]{Shekhter13}%
  \BibitemOpen
  \bibfield  {author} {\bibinfo {author} {\bibfnamefont {A.}~\bibnamefont
  {Shekhter}}, \bibinfo {author} {\bibfnamefont {B.~J.}\ \bibnamefont
  {Ramshaw}}, \bibinfo {author} {\bibfnamefont {R.}~\bibnamefont {Liang}},
  \bibinfo {author} {\bibfnamefont {W.~N.}\ \bibnamefont {Hardy}}, \bibinfo
  {author} {\bibfnamefont {D.~A.}\ \bibnamefont {Bonn}}, \bibinfo {author}
  {\bibfnamefont {F.~F.}\ \bibnamefont {Balakirev}}, \bibinfo {author}
  {\bibfnamefont {R.~D.}\ \bibnamefont {McDonald}}, \bibinfo {author}
  {\bibfnamefont {J.~B.}\ \bibnamefont {Betts}}, \bibinfo {author}
  {\bibfnamefont {S.~C.}\ \bibnamefont {Riggs}}, \ and\ \bibinfo {author}
  {\bibfnamefont {A.}~\bibnamefont {Migliori}},\ }\href {\doibase
  10.1038/nature12165} {\bibfield  {journal} {\bibinfo  {journal} {Nature}\
  }\textbf {\bibinfo {volume} {498}},\ \bibinfo {pages} {75} (\bibinfo {year}
  {2013})}\BibitemShut {NoStop}%
\bibitem [{\citenamefont {{Orth}}\ \emph {et~al.}(2017)\citenamefont {{Orth}},
  \citenamefont {{Jeevanesan}}, \citenamefont {{Fernandes}},\ and\
  \citenamefont {{Schmalian}}}]{Orth2017}%
  \BibitemOpen
  \bibfield  {author} {\bibinfo {author} {\bibfnamefont {P.~P.}\ \bibnamefont
  {{Orth}}}, \bibinfo {author} {\bibfnamefont {B.}~\bibnamefont
  {{Jeevanesan}}}, \bibinfo {author} {\bibfnamefont {R.~M.}\ \bibnamefont
  {{Fernandes}}}, \ and\ \bibinfo {author} {\bibfnamefont {J.}~\bibnamefont
  {{Schmalian}}},\ }\href@noop {} {\  (\bibinfo {year} {2017})},\ \Eprint
  {http://arxiv.org/abs/1703.02210} {arXiv:1703.02210} \BibitemShut {NoStop}%
\bibitem [{\citenamefont {Agterberg}\ \emph {et~al.}(2015)\citenamefont
  {Agterberg}, \citenamefont {Melchert},\ and\ \citenamefont
  {Kashyap}}]{Agterberg:2014wf}%
  \BibitemOpen
  \bibfield  {author} {\bibinfo {author} {\bibfnamefont {D.~F.}\ \bibnamefont
  {Agterberg}}, \bibinfo {author} {\bibfnamefont {D.~S.}\ \bibnamefont
  {Melchert}}, \ and\ \bibinfo {author} {\bibfnamefont {M.~K.}\ \bibnamefont
  {Kashyap}},\ }\href {\doibase 10.1103/PhysRevB.91.054502} {\bibfield
  {journal} {\bibinfo  {journal} {Phys. Rev. B}\ }\textbf {\bibinfo {volume}
  {91}},\ \bibinfo {pages} {054502} (\bibinfo {year} {2015})}\BibitemShut
  {NoStop}%
\bibitem [{\citenamefont {Einenkel}\ \emph {et~al.}(2014)\citenamefont
  {Einenkel}, \citenamefont {Meier}, \citenamefont {P\'epin},\ and\
  \citenamefont {Efetov}}]{Einenkel14}%
  \BibitemOpen
  \bibfield  {author} {\bibinfo {author} {\bibfnamefont {M.}~\bibnamefont
  {Einenkel}}, \bibinfo {author} {\bibfnamefont {H.}~\bibnamefont {Meier}},
  \bibinfo {author} {\bibfnamefont {C.}~\bibnamefont {P\'epin}}, \ and\
  \bibinfo {author} {\bibfnamefont {K.~B.}\ \bibnamefont {Efetov}},\ }\href
  {\doibase 10.1103/PhysRevB.90.054511} {\bibfield  {journal} {\bibinfo
  {journal} {Phys. Rev. B}\ }\textbf {\bibinfo {volume} {90}},\ \bibinfo
  {pages} {054511} (\bibinfo {year} {2014})}\BibitemShut {NoStop}%
\bibitem [{\citenamefont {Kohsaka}\ \emph {et~al.}(2008)\citenamefont
  {Kohsaka}, \citenamefont {Taylor}, \citenamefont {Wahl}, \citenamefont
  {Schmidt}, \citenamefont {Lee},\ and\ \citenamefont
  {Fujita}}]{Kohsaka:2008bz}%
  \BibitemOpen
  \bibfield  {author} {\bibinfo {author} {\bibfnamefont {Y.}~\bibnamefont
  {Kohsaka}}, \bibinfo {author} {\bibfnamefont {C.}~\bibnamefont {Taylor}},
  \bibinfo {author} {\bibfnamefont {P.}~\bibnamefont {Wahl}}, \bibinfo {author}
  {\bibfnamefont {A.}~\bibnamefont {Schmidt}}, \bibinfo {author} {\bibfnamefont
  {J.}~\bibnamefont {Lee}}, \ and\ \bibinfo {author} {\bibfnamefont
  {K.}~\bibnamefont {Fujita}},\ }\href {\doibase 10.1038/nature07243}
  {\bibfield  {journal} {\bibinfo  {journal} {Nature}\ }\textbf {\bibinfo
  {volume} {454}},\ \bibinfo {pages} {1072} (\bibinfo {year}
  {2008})}\BibitemShut {NoStop}%
\bibitem [{\citenamefont {Wu}\ \emph {et~al.}(2011)\citenamefont {Wu},
  \citenamefont {Mayaffre}, \citenamefont {Kr{\"a}mer}, \citenamefont
  {Horvatic}, \citenamefont {Berthier}, \citenamefont {Hardy}, \citenamefont
  {Liang}, \citenamefont {Bonn},\ and\ \citenamefont {Julien}}]{Wu11}%
  \BibitemOpen
  \bibfield  {author} {\bibinfo {author} {\bibfnamefont {T.}~\bibnamefont
  {Wu}}, \bibinfo {author} {\bibfnamefont {H.}~\bibnamefont {Mayaffre}},
  \bibinfo {author} {\bibfnamefont {S.}~\bibnamefont {Kr{\"a}mer}}, \bibinfo
  {author} {\bibfnamefont {M.}~\bibnamefont {Horvatic}}, \bibinfo {author}
  {\bibfnamefont {C.}~\bibnamefont {Berthier}}, \bibinfo {author}
  {\bibfnamefont {W.~N.}\ \bibnamefont {Hardy}}, \bibinfo {author}
  {\bibfnamefont {R.}~\bibnamefont {Liang}}, \bibinfo {author} {\bibfnamefont
  {D.~A.}\ \bibnamefont {Bonn}}, \ and\ \bibinfo {author} {\bibfnamefont
  {M.-H.}\ \bibnamefont {Julien}},\ }\href {\doibase 10.1038/nature10345}
  {\bibfield  {journal} {\bibinfo  {journal} {Nature}\ }\textbf {\bibinfo
  {volume} {477}},\ \bibinfo {pages} {191} (\bibinfo {year}
  {2011})}\BibitemShut {NoStop}%
\bibitem [{\citenamefont {LeBoeuf}\ \emph {et~al.}(2013)\citenamefont
  {LeBoeuf}, \citenamefont {Kramer}, \citenamefont {Hardy}, \citenamefont
  {Liang}, \citenamefont {Bonn},\ and\ \citenamefont {Proust}}]{LeBoeuf13}%
  \BibitemOpen
  \bibfield  {author} {\bibinfo {author} {\bibfnamefont {D.}~\bibnamefont
  {LeBoeuf}}, \bibinfo {author} {\bibfnamefont {S.}~\bibnamefont {Kramer}},
  \bibinfo {author} {\bibfnamefont {W.~N.}\ \bibnamefont {Hardy}}, \bibinfo
  {author} {\bibfnamefont {R.}~\bibnamefont {Liang}}, \bibinfo {author}
  {\bibfnamefont {D.~A.}\ \bibnamefont {Bonn}}, \ and\ \bibinfo {author}
  {\bibfnamefont {C.}~\bibnamefont {Proust}},\ }\href
  {http://dx.doi.org/10.1038/nphys2502} {\bibfield  {journal} {\bibinfo
  {journal} {Nat. Phys.}\ }\textbf {\bibinfo {volume} {9}},\ \bibinfo {pages}
  {79} (\bibinfo {year} {2013})}\BibitemShut {NoStop}%
\bibitem [{\citenamefont {Yu}\ \emph {et~al.}(2016)\citenamefont {Yu},
  \citenamefont {Hirschberger}, \citenamefont {Loew}, \citenamefont {Li},
  \citenamefont {Lawson}, \citenamefont {Asaba}, \citenamefont {Kemper},
  \citenamefont {Liang}, \citenamefont {Porras}, \citenamefont {Boebinger},
  \citenamefont {Singleton}, \citenamefont {Keimer}, \citenamefont {Li},\ and\
  \citenamefont {Ong}}]{Yu12667}%
  \BibitemOpen
  \bibfield  {author} {\bibinfo {author} {\bibfnamefont {F.}~\bibnamefont
  {Yu}}, \bibinfo {author} {\bibfnamefont {M.}~\bibnamefont {Hirschberger}},
  \bibinfo {author} {\bibfnamefont {T.}~\bibnamefont {Loew}}, \bibinfo {author}
  {\bibfnamefont {G.}~\bibnamefont {Li}}, \bibinfo {author} {\bibfnamefont
  {B.~J.}\ \bibnamefont {Lawson}}, \bibinfo {author} {\bibfnamefont
  {T.}~\bibnamefont {Asaba}}, \bibinfo {author} {\bibfnamefont {J.~B.}\
  \bibnamefont {Kemper}}, \bibinfo {author} {\bibfnamefont {T.}~\bibnamefont
  {Liang}}, \bibinfo {author} {\bibfnamefont {J.}~\bibnamefont {Porras}},
  \bibinfo {author} {\bibfnamefont {G.~S.}\ \bibnamefont {Boebinger}}, \bibinfo
  {author} {\bibfnamefont {J.}~\bibnamefont {Singleton}}, \bibinfo {author}
  {\bibfnamefont {B.}~\bibnamefont {Keimer}}, \bibinfo {author} {\bibfnamefont
  {L.}~\bibnamefont {Li}}, \ and\ \bibinfo {author} {\bibfnamefont {N.~P.}\
  \bibnamefont {Ong}},\ }\href {\doibase 10.1073/pnas.1612591113} {\bibfield
  {journal} {\bibinfo  {journal} {Proc. Nat. Ac. Sci.}\ }\textbf {\bibinfo
  {volume} {113}},\ \bibinfo {pages} {12667} (\bibinfo {year}
  {2016})}\BibitemShut {NoStop}%
\bibitem [{\citenamefont {Gerber}\ \emph {et~al.}(2015)\citenamefont {Gerber},
  \citenamefont {Jang}, \citenamefont {Nojiri}, \citenamefont {Matsuzawa},
  \citenamefont {Yasumura}, \citenamefont {Bonn}, \citenamefont {Liang},
  \citenamefont {Hardy}, \citenamefont {Islam}, \citenamefont {Mehta},
  \citenamefont {Song}, \citenamefont {Sikorski}, \citenamefont {Stefanescu},
  \citenamefont {Feng}, \citenamefont {Kivelson}, \citenamefont {Devereaux},
  \citenamefont {Shen}, \citenamefont {Kao}, \citenamefont {Lee}, \citenamefont
  {Zhu},\ and\ \citenamefont {Lee}}]{Gerber:2015gx}%
  \BibitemOpen
  \bibfield  {author} {\bibinfo {author} {\bibfnamefont {S.}~\bibnamefont
  {Gerber}}, \bibinfo {author} {\bibfnamefont {H.}~\bibnamefont {Jang}},
  \bibinfo {author} {\bibfnamefont {H.}~\bibnamefont {Nojiri}}, \bibinfo
  {author} {\bibfnamefont {S.}~\bibnamefont {Matsuzawa}}, \bibinfo {author}
  {\bibfnamefont {H.}~\bibnamefont {Yasumura}}, \bibinfo {author}
  {\bibfnamefont {D.~A.}\ \bibnamefont {Bonn}}, \bibinfo {author}
  {\bibfnamefont {R.}~\bibnamefont {Liang}}, \bibinfo {author} {\bibfnamefont
  {W.~N.}\ \bibnamefont {Hardy}}, \bibinfo {author} {\bibfnamefont
  {Z.}~\bibnamefont {Islam}}, \bibinfo {author} {\bibfnamefont
  {A.}~\bibnamefont {Mehta}}, \bibinfo {author} {\bibfnamefont
  {S.}~\bibnamefont {Song}}, \bibinfo {author} {\bibfnamefont {M.}~\bibnamefont
  {Sikorski}}, \bibinfo {author} {\bibfnamefont {D.}~\bibnamefont
  {Stefanescu}}, \bibinfo {author} {\bibfnamefont {Y.}~\bibnamefont {Feng}},
  \bibinfo {author} {\bibfnamefont {S.~A.}\ \bibnamefont {Kivelson}}, \bibinfo
  {author} {\bibfnamefont {T.~P.}\ \bibnamefont {Devereaux}}, \bibinfo {author}
  {\bibfnamefont {Z.-X.}\ \bibnamefont {Shen}}, \bibinfo {author}
  {\bibfnamefont {C.~C.}\ \bibnamefont {Kao}}, \bibinfo {author} {\bibfnamefont
  {W.~S.}\ \bibnamefont {Lee}}, \bibinfo {author} {\bibfnamefont
  {D.}~\bibnamefont {Zhu}}, \ and\ \bibinfo {author} {\bibfnamefont {J.~S.}\
  \bibnamefont {Lee}},\ }\href {\doibase 10.1126/science.aac6257} {\bibfield
  {journal} {\bibinfo  {journal} {Science}\ }\textbf {\bibinfo {volume}
  {350}},\ \bibinfo {pages} {949} (\bibinfo {year} {2015})}\BibitemShut
  {NoStop}%
\bibitem [{\citenamefont {Chang}\ \emph {et~al.}(2016)\citenamefont {Chang},
  \citenamefont {Blackburn}, \citenamefont {Ivashko}, \citenamefont {Holmes},
  \citenamefont {Christensen}, \citenamefont {Huecker}, \citenamefont {Liang},
  \citenamefont {Bonn}, \citenamefont {Hardy}, \citenamefont {Ruett},
  \citenamefont {Zimmermann}, \citenamefont {Forgan},\ and\ \citenamefont
  {Hayden}}]{Chang:2016gz}%
  \BibitemOpen
  \bibfield  {author} {\bibinfo {author} {\bibfnamefont {J.}~\bibnamefont
  {Chang}}, \bibinfo {author} {\bibfnamefont {E.}~\bibnamefont {Blackburn}},
  \bibinfo {author} {\bibfnamefont {O.}~\bibnamefont {Ivashko}}, \bibinfo
  {author} {\bibfnamefont {A.~T.}\ \bibnamefont {Holmes}}, \bibinfo {author}
  {\bibfnamefont {N.~B.}\ \bibnamefont {Christensen}}, \bibinfo {author}
  {\bibfnamefont {M.}~\bibnamefont {Huecker}}, \bibinfo {author} {\bibfnamefont
  {R.}~\bibnamefont {Liang}}, \bibinfo {author} {\bibfnamefont {D.~A.}\
  \bibnamefont {Bonn}}, \bibinfo {author} {\bibfnamefont {W.~N.}\ \bibnamefont
  {Hardy}}, \bibinfo {author} {\bibfnamefont {U.}~\bibnamefont {Ruett}},
  \bibinfo {author} {\bibfnamefont {M.~V.}\ \bibnamefont {Zimmermann}},
  \bibinfo {author} {\bibfnamefont {E.~M.}\ \bibnamefont {Forgan}}, \ and\
  \bibinfo {author} {\bibfnamefont {S.~M.}\ \bibnamefont {Hayden}},\ }\href
  {\doibase 10.1038/ncomms11494} {\bibfield  {journal} {\bibinfo  {journal}
  {Nat. Commun.}\ }\textbf {\bibinfo {volume} {7}} (\bibinfo {year} {2016}),\
  10.1038/ncomms11494}\BibitemShut {NoStop}%
\bibitem [{\citenamefont {Zhang}(1997)}]{Zhang:1997ew}%
  \BibitemOpen
  \bibfield  {author} {\bibinfo {author} {\bibfnamefont {S.~C.}\ \bibnamefont
  {Zhang}},\ }\href {\doibase 10.1126/science.275.5303.1089} {\bibfield
  {journal} {\bibinfo  {journal} {Science}\ }\textbf {\bibinfo {volume}
  {275}},\ \bibinfo {pages} {1089} (\bibinfo {year} {1997})}\BibitemShut
  {NoStop}%
\bibitem [{\citenamefont {Chakravarty}\ \emph {et~al.}(1988)\citenamefont
  {Chakravarty}, \citenamefont {Halperin},\ and\ \citenamefont
  {Nelson}}]{Chakravarty:1988uu}%
  \BibitemOpen
  \bibfield  {author} {\bibinfo {author} {\bibfnamefont {S.}~\bibnamefont
  {Chakravarty}}, \bibinfo {author} {\bibfnamefont {B.}~\bibnamefont
  {Halperin}}, \ and\ \bibinfo {author} {\bibfnamefont {D.}~\bibnamefont
  {Nelson}},\ }\href@noop {} {\bibfield  {journal} {\bibinfo  {journal} {Phys.
  Rev. Lett.}\ }\textbf {\bibinfo {volume} {60}} (\bibinfo {year}
  {1988})}\BibitemShut {NoStop}%
\bibitem [{\citenamefont {Rosenstein}\ and\ \citenamefont
  {Li}(2010)}]{Rosenstein2010}%
  \BibitemOpen
  \bibfield  {author} {\bibinfo {author} {\bibfnamefont {B.}~\bibnamefont
  {Rosenstein}}\ and\ \bibinfo {author} {\bibfnamefont {D.}~\bibnamefont
  {Li}},\ }\href {\doibase 10.1103/RevModPhys.82.109} {\bibfield  {journal}
  {\bibinfo  {journal} {Rev. Mod. Phys.}\ }\textbf {\bibinfo {volume} {82}},\
  \bibinfo {pages} {109} (\bibinfo {year} {2010})}\BibitemShut {NoStop}%
\bibitem [{\citenamefont {Yang}(1989)}]{Yang89}%
  \BibitemOpen
  \bibfield  {author} {\bibinfo {author} {\bibfnamefont {C.~N.}\ \bibnamefont
  {Yang}},\ }\href {\doibase 10.1103/PhysRevLett.63.2144} {\bibfield  {journal}
  {\bibinfo  {journal} {Phys. Rev. Lett.}\ }\textbf {\bibinfo {volume} {63}},\
  \bibinfo {pages} {2144} (\bibinfo {year} {1989})}\BibitemShut {NoStop}%
\bibitem [{\citenamefont {Yang}\ and\ \citenamefont
  {Zhang}(1990)}]{Yang:1990cf}%
  \BibitemOpen
  \bibfield  {author} {\bibinfo {author} {\bibfnamefont {C.~N.}\ \bibnamefont
  {Yang}}\ and\ \bibinfo {author} {\bibfnamefont {S.~C.}\ \bibnamefont
  {Zhang}},\ }\href {\doibase 10.1142/S0217984990000933} {\bibfield  {journal}
  {\bibinfo  {journal} {Mod. Phys. Lett. B}\ }\textbf {\bibinfo {volume}
  {04}},\ \bibinfo {pages} {759} (\bibinfo {year} {1990})}\BibitemShut
  {NoStop}%
\bibitem [{\citenamefont {Zhang}(1990)}]{Zhang1990}%
  \BibitemOpen
  \bibfield  {author} {\bibinfo {author} {\bibfnamefont {S.}~\bibnamefont
  {Zhang}},\ }\href {\doibase 10.1103/PhysRevLett.65.120} {\bibfield  {journal}
  {\bibinfo  {journal} {Phys. Rev. Lett.}\ }\textbf {\bibinfo {volume} {65}},\
  \bibinfo {pages} {120} (\bibinfo {year} {1990})}\BibitemShut {NoStop}%
\bibitem [{\citenamefont {Davis}(2017)}]{DavisPrivate}%
  \BibitemOpen
  \bibfield  {author} {\bibinfo {author} {\bibfnamefont {J.~C.}\ \bibnamefont
  {Davis}},\ }\href@noop {} {\bibfield  {journal} {\bibinfo  {journal} {Private
  communication}\ } (\bibinfo {year} {2017})}\BibitemShut {NoStop}%
\bibitem [{Gri(2014)}]{Grissonnanche14}%
  \BibitemOpen
  \href {\doibase 10.1038/ncomms4280} {\bibfield  {journal} {\bibinfo
  {journal} {Nat. Commun.}\ }\textbf {\bibinfo {volume} {5}} (\bibinfo {year}
  {2014}),\ 10.1038/ncomms4280}\BibitemShut {NoStop}%
\bibitem [{\citenamefont {Hsu}\ \emph {et~al.}(2017)\citenamefont {Hsu},
  \citenamefont {Hartstein}, \citenamefont {Davies}, \citenamefont {Chan},
  \citenamefont {Porras}, \citenamefont {Loew}, \citenamefont {Taylor},
  \citenamefont {Liu}, \citenamefont {Tacon}, \citenamefont {Zuo},
  \citenamefont {Wang}, \citenamefont {Zhu}, \citenamefont {Lonzarich},
  \citenamefont {Keimer}, \citenamefont {Harrison},\ and\ \citenamefont
  {Sebastian}}]{Hsu2017}%
  \BibitemOpen
  \bibfield  {author} {\bibinfo {author} {\bibfnamefont {Y.-T.}\ \bibnamefont
  {Hsu}}, \bibinfo {author} {\bibfnamefont {M.}~\bibnamefont {Hartstein}},
  \bibinfo {author} {\bibfnamefont {A.~J.}\ \bibnamefont {Davies}}, \bibinfo
  {author} {\bibfnamefont {M.~K.}\ \bibnamefont {Chan}}, \bibinfo {author}
  {\bibfnamefont {J.}~\bibnamefont {Porras}}, \bibinfo {author} {\bibfnamefont
  {T.}~\bibnamefont {Loew}}, \bibinfo {author} {\bibfnamefont {S.~V.}\
  \bibnamefont {Taylor}}, \bibinfo {author} {\bibfnamefont {H.}~\bibnamefont
  {Liu}}, \bibinfo {author} {\bibfnamefont {M.~L.}\ \bibnamefont {Tacon}},
  \bibinfo {author} {\bibfnamefont {H.}~\bibnamefont {Zuo}}, \bibinfo {author}
  {\bibfnamefont {J.}~\bibnamefont {Wang}}, \bibinfo {author} {\bibfnamefont
  {Z.}~\bibnamefont {Zhu}}, \bibinfo {author} {\bibfnamefont {G.~G.}\
  \bibnamefont {Lonzarich}}, \bibinfo {author} {\bibfnamefont {B.}~\bibnamefont
  {Keimer}}, \bibinfo {author} {\bibfnamefont {N.}~\bibnamefont {Harrison}}, \
  and\ \bibinfo {author} {\bibfnamefont {S.~E.}\ \bibnamefont {Sebastian}},\
  }\href@noop {} {\bibfield  {journal} {\bibinfo  {journal} {in preparation}\ }
  (\bibinfo {year} {2017})}\BibitemShut {NoStop}%
\bibitem [{\citenamefont {Sigrist}\ and\ \citenamefont
  {Ueda}(1991)}]{Sigrist1991}%
  \BibitemOpen
  \bibfield  {author} {\bibinfo {author} {\bibfnamefont {M.}~\bibnamefont
  {Sigrist}}\ and\ \bibinfo {author} {\bibfnamefont {K.}~\bibnamefont {Ueda}},\
  }\href {\doibase 10.1103/RevModPhys.63.239} {\bibfield  {journal} {\bibinfo
  {journal} {Rev. Mod. Phys.}\ }\textbf {\bibinfo {volume} {63}},\ \bibinfo
  {pages} {239} (\bibinfo {year} {1991})}\BibitemShut {NoStop}%
\end{thebibliography}%

%%%%%%%%%% Merge with supplemental materials %%%%%%%%%%
\widetext
\clearpage
\begin{center}
\textbf{\large Supplemental material for:\\Pseudo-spin Skyrmions in the Phase Diagram of Cuprate Superconductors}
\end{center}
%%%%%%%%%% Merge with supplemental materials %%%%%%%%%%
%%%%%%%%%% Prefix a "S" to all equations, figures, tables and reset the counter %%%%%%%%%%
\setcounter{equation}{0}
\setcounter{figure}{0}
\setcounter{table}{0}
\setcounter{page}{1}
\makeatletter
\renewcommand{\theequation}{S\arabic{equation}}
\renewcommand{\thefigure}{S\arabic{figure}}
\renewcommand{\bibnumfmt}[1]{[S#1]}
%\renewcommand{\citenumfont}[1]{S#1}
%%%%%%%%%% Prefix a "S" to all equations, figures, tables and reset the counter %%%%%%%%%%

\section*{Loop currents}

Here, for convenience, we give the details of the calculation of the loop current order parameter defined in the main text under the time-reversal ($\mathcal{T}$) and parity ($\mathcal{P}$) transformations. It closely follows the argument given in \cite{Agterberg:2014wf}. We start by considering the charge order parameter, which was derived from the $t$-$J$ model in a previous work (\cite{Montiel16} Eq.\ 7b):
\begin{equation}
\chi_{\textbf{Q}_0} = \frac{1}{2} \sum_{\textbf{k},\sigma} \left[ \cos (2 \theta_\textbf{k}) + \cos (2 \theta_{\bar{\textbf{k}}} ) \right] \psi^\dagger_{\bar{\textbf{k}},\sigma} \psi_{-\textbf{k},\sigma}
\end{equation}
where $\theta_{\textbf{k}}$ is the angle spanning the Brillouin zone and $\bar{\textbf{k}}$ is the involution:
\begin{equation}
\bar{\textbf{k}} = -\textbf{k} + \textbf{Q}_0
\end{equation}
where $\textbf{Q}_0$ is the ordering wave vector of the charge order parameter defined in the main text.

The time-reversal operation acts on annihilation operators following \cite{Sigrist1991}:
\begin{equation}
\mathcal{T} \left( \psi_{\textbf{k},\sigma} \right) = \sum_{\sigma '} \left( -i \bf{\sigma}^y \right)_{\sigma \sigma '} \psi^\dagger_{-\textbf{k},\sigma'}
\end{equation}
where $\sigma$ and $\sigma'$ are spin indices, and $\sigma^y$ is a Pauli matrix. This can be written more simply as:
\begin{equation}
\mathcal{T} \left( \psi_{\textbf{k},\sigma} \right) = -sgn(\sigma) \psi^\dagger_{-\textbf{k},-\sigma}
\end{equation}
where $sgn(\uparrow) = 1$ and $sgn(\downarrow) = -1$. Applying time-reversal to the charge order parameter gives:
\begin{align}
\mathcal{T} \left( \chi_{\textbf{Q}_0} \right) = &-\frac{1}{2} \sum_{\textbf{k},\sigma} \left[ \cos (2 \theta_{\textbf{k}}) + \cos (2 \theta_{-\textbf{k} + \textbf{Q}_0} ) \right] sgn(\sigma)^2 \psi^\dagger_{\textbf{k},-\sigma} \psi_{\textbf{k} - \textbf{Q}_0,-\sigma} 
\\
= &-\frac{1}{2} \sum_{\textbf{k}',\sigma'} \left[ \cos (2 \theta_{-\textbf{k}'}) + \cos (2 \theta_{\textbf{k}' + \textbf{Q}_0} ) \right] \psi^\dagger_{-\textbf{k}',\sigma'} \psi_{-\textbf{k}' - \textbf{Q}_0,\sigma'} 
\\
= &\frac{1}{2} \sum_{\textbf{k}',\sigma'} \left[ \cos (2 \theta_{\textbf{k}'}) + \cos (2 \theta_{-\textbf{k}' - \textbf{Q}_0} ) \right] \left( \psi^\dagger_{-\textbf{k}' - \textbf{Q}_0,\sigma'} \psi_{-\textbf{k}',\sigma'} \right)^\dagger
\\
= & \chi^\dagger_{-\textbf{Q}_0}
\end{align}
Let us now consider the $d$-wave superconducting order parameter:
\begin{equation}
\Delta_{\textbf{Q}_0}^\dagger = \frac{1}{\sqrt{2}} \sum_{\textbf{k}, \sigma}  \left[ \cos (2 \theta_{\textbf{k}}) + \cos (2 \theta_{-\textbf{k} + \textbf{Q}_0} ) \right] \psi^\dagger_{\textbf{k},\sigma} \psi^\dagger_{-\textbf{k},-\sigma}
\end{equation}
Here we obtain:
\begin{align}
\mathcal{T} \left( \Delta_{\textbf{Q}_0}^\dagger \right) = &\frac{1}{\sqrt{2}} \sum_{\textbf{k}, \sigma}  \left[ \cos (2 \theta_{\textbf{k}}) + \cos (2 \theta_{-\textbf{k} + \textbf{Q}_0} ) \right] \psi_{\textbf{k},\sigma} \left( -\psi_{-\textbf{k},-\sigma} \right)
\\
= &-\frac{1}{\sqrt{2}} \sum_{\textbf{k}', \sigma'}  \left[ \cos (2 \theta_{-\textbf{k}'}) + \cos (2 \theta_{\textbf{k}' + \textbf{Q}_0} ) \right] \psi_{-\textbf{k}',-\sigma'} \psi_{\textbf{k}',\sigma'}
\\
= &\frac{1}{\sqrt{2}} \sum_{\textbf{k}', \sigma'}  \left[ \cos (2 \theta_{\textbf{k}'}) + \cos (2 \theta_{-\textbf{k}' - \textbf{Q}_0} ) \right] \left( \psi_{\textbf{k}',\sigma'} \psi_{-\textbf{k}',-\sigma'} \right)^\dagger
\\
= &\Delta_{-\textbf{Q}_0}
\end{align}

Note that here the operator $\Delta$ is not Hermitian. Gathering these two results gives:
\begin{equation}
\mathcal{T}(\phi_{\textbf{Q}_0}) = \mathcal{T} (\chi_{\textbf{Q}_0 }\Delta_{\textbf{Q}_0}) = \Delta^\dagger_{-\textbf{Q}_0} \chi^\dagger_{-\textbf{Q}_0} = \phi^\dagger_{-\textbf{Q}_0}
\end{equation}
The case of parity is much simpler and yields:
\begin{align}
\mathcal{P} \left( \phi_{\textbf{Q}_0} \right) = \phi_{-\textbf{Q}_0}
\end{align}
We now define the loop current order parameter as:
\begin{equation}
l = |\phi_{\textbf{Q}_0}|^2 - |\phi_{-\textbf{Q}_0}|^2
\end{equation}
which therefore transforms as:
\begin{equation}
l \xrightarrow[]{\mathcal{T}} -l, \quad l \xrightarrow[]{\mathcal{P}} -l, \quad l \xrightarrow[]{\mathcal{TP}} l
\end{equation}

\end{document}